\documentclass[
article,superscriptaddress,preprint]{revtex4-2}

\preprint{APS/123-QED}
\usepackage{amsmath}
\usepackage{graphicx}
\usepackage{bm}
\usepackage{mathrsfs}
\usepackage{lineno}
\usepackage{setspace}
\usepackage{float}
\usepackage{comment}
\usepackage{scalerel}
\begin{document}

\title{Exact solutions for the electromagnetic fields of a flying focus}



\author{D. Ramsey}
\email{dram@lle.rochester.edu}
\affiliation{
University of Rochester, Laboratory for Laser Energetics, Rochester, New York, 14623-1299 USA
}
\author{A. Di Piazza}
\affiliation{
Max-Planck-Institut für Kernphysik, Saupfercheckweg 1, D-69117 Heidelberg, Germany
}
\author{M. Formanek}
\affiliation{
ELI-Beamlines, Institute of Physics of Czech Academy of Sciences, Dolní Břežany 252 41, Czech Republic
}
\author{P. Franke}
\affiliation{
University of Rochester, Laboratory for Laser Energetics, Rochester, New York, 14623-1299 USA
}
\author{D.H. Froula}
\affiliation{
University of Rochester, Laboratory for Laser Energetics, Rochester, New York, 14623-1299 USA
}
\author{B. Malaca}
\affiliation{
GoLP/Instituto de Plasmas e Fusão Nuclear, Instituto Superior Técnico, Universidade de Lisboa, Lisbon 1049-001, Portugal
}
\author{W.B. Mori}
\affiliation{
University of California, Los Angeles, California 90095, USA
}
\author{J.R. Pierce}
\affiliation{
University of California, Los Angeles, California 90095, USA
}
\author{T.T. Simpson}
\affiliation{
University of Rochester, Laboratory for Laser Energetics, Rochester, New York, 14623-1299 USA
}
\author{J. Vieira}
\affiliation{
GoLP/Instituto de Plasmas e Fusão Nuclear, Instituto Superior Técnico, Universidade de Lisboa, Lisbon 1049-001, Portugal
}
\author{M. Vranic}
\affiliation{
GoLP/Instituto de Plasmas e Fusão Nuclear, Instituto Superior Técnico, Universidade de Lisboa, Lisbon 1049-001, Portugal
}
\author{K. Weichman}
\affiliation{
University of Rochester, Laboratory for Laser Energetics, Rochester, New York, 14623-1299 USA
}
\author{J.P. Palastro} 
\email{jpal@lle.rochester.edu}
\affiliation{
University of Rochester, Laboratory for Laser Energetics, Rochester, New York, 14623-1299 USA
}

\date{\today}
\newcommand{\partialD}[1]{\ensuremath{\frac{\partial}{\partial #1}} }    
\newcommand{\partialDsq}[1]{\ensuremath{\frac{\partial^2}{\partial #1^2}} } 
\newcommand{\cpx}{\ensuremath{\tfrac{1}{2}i\kappa_0 w_0^2}}
\newcommand{\unitvec}[1]{\ensuremath{\bm{\hat{\mathrm{#1}}}}}

\newcommand{\bvec}[1]{\ensuremath{\bm{\mathrm{#1}}}}

\begin{abstract}
The intensity peak of a "flying" focus travels at a programmable velocity over many Rayleigh ranges while maintaining a near-constant profile. Assessing the extent to which these features can enhance laser-based applications requires an accurate description of the electromagnetic fields. Here we present exact analytical solutions to Maxwell’s equations for the electromagnetic fields of a constant-velocity flying focus, generalized for arbitrary polarization and orbital angular momentum. The approach combines the complex source-point method, which transforms multipole solutions into beam-like solutions, with the Lorentz invariance of Maxwell's equations. Propagating the fields backward in space reveals the space-time profile that an optical assembly must produce to realize these fields in the laboratory. Comparisons with simpler paraxial solutions provide conditions for their reliable use when modeling a flying focus.
\end{abstract}

\maketitle



\section{Introduction}
All focused laser fields exhibit a moving focus in some frame of reference. In the laboratory frame, an ideal lens focuses every frequency, temporal slice, and annulus of a laser pulse to the same location. The pulse moves through the focus at its group velocity and diffracts over a Rayleigh range. In any other Lorentz frame, the focus moves. ``Flying focus'' techniques recreate these moving foci in the laboratory frame by modifying the focal time and location of each frequency, temporal slice, or annulus of a pulse \cite{sainte2017controlling,froula2018spatiotemporal,simpson2020nonlinear,simpson2022spatiotemporal,palastro2020dephasingless,pierce2022ASTRL}. The intensity peak formed by the moving focus can travel at any arbitrary velocity while maintaining a near-constant profile over many Rayleigh ranges. 

The first experimental demonstration of a flying focus used chromatic focusing of a chirped laser pulse to control the focal time and location of each frequency \cite{froula2018spatiotemporal}. This technique, referred to as the ``chromatic'' flying focus, limits the bandwidth available at each focal location, which places a lower bound on the duration of the intensity peak. To address the need for ultrashort intensity peaks, two alternative techniques have been proposed. The ``flying focus X'' uses cross-phase modulation in a Kerr lens to imprint a different focusing phase onto each temporal slice of a pulse  \cite{simpson2022spatiotemporal}. The time-dependent refractive index experienced by the pulse in the Kerr lens provides the bandwidth necessary to support the duration of the intensity peak.  The ``achromatic'' flying focus combines an axiparabola \cite{smartsev2019axiparabola} with a radial echelon to control the focal location and relative timing of each annulus, respectively \cite{palastro2020dephasingless}. As the annuli come in and out of focus, they interfere to form an intensity peak with a duration equal to that of the initial pulse. 

The programmable velocity $\text{v}_I$ and extended focal range $L$ of a flying focus offer new approaches to realizing or optimizing laser-based applications. The intensity peak of a flying focus pulse can travel slower than the group velocity; faster than the group velocity, i.e., superluminally; or backward with respect to the phase fronts of the pulse. Superluminal intensity peaks have been proposed to overcome dephasing and wave breaking in laser wakefield acceleration \cite{palastro2020dephasingless,Debus2019,Caizergues2020,palastro2021laser} and to increase the rate of frequency upshifting in photon acceleration \cite{franke2021optical}. Backward intensity peaks can facilitate the formation of long plasma channels by mitigating ionization refraction \cite{palastro2018ionization,howard2019photon} and may improve the performance of Raman amplifiers by ensuring quasi-stationary plasma conditions \cite{Turnbull2018}. The motion of a backward intensity peak against its phase fronts also allows for longer interaction lengths in fundamental studies of nonlinear Compton scattering and radiation reaction, which can amplify observable signatures of these processes \cite{di2021scattering,formanek2022radiation}. Further, a backward intensity peak can ponderomotively accelerate electrons to relativistic momenta in the backward direction, providing unprecedented control over the electron trajectory and greatly enhancing the radiation properties in nonlinear Thomson scattering \cite{ramsey2020vacuum,ramsey2022nonlinear}. 

Assessing the extent to which a flying focus can enable or enhance these applications requires an accurate description of the electromagnetic fields. With the exception of the special case $\text{v}_I = -c$ \cite{di2021scattering,formanek2022radiation}, all of the aforementioned applications were modeled using approximate solutions for the electromagnetic fields of flying focus pulses. In the case of conventional pulses with stationary foci, improving the accuracy of approximate solutions has been found to impact models of phenomena ranging from direct laser acceleration to optical trapping \cite{barton1989theoretical,neuman2004optical,cicchitelli1990longitudinal,esarey1995theory,esarey1995laser,quesnel1998theory,hora2000principle}. Methods for obtaining accurate solutions to Maxwell's equations for conventional laser pulses come in three forms: a ``Lax''-like series expansion in which corrections to paraxial fields can be calculated recursively \cite{lax1975maxwell,davis1979theory,barton1989fifth,salamin2007fields}; series expansions of exact spectral integrals for each field component \cite{agrawal1979gaussian,cicchitelli1990longitudinal,quesnel1998theory}; and the complex source-point method (CSPM), which exploits the invariance of Maxwell's equations under a translation in the complex plane to transform multipole solutions into beam-like solutions \cite{deschamps1971gaussian,shin1977gaussian, zauderer1986complex, norris1986complex,heyman2001gaussian,cullen1979complex,sheppard1999electromagnetic,mitri2013quasi}. Of these three, the CSPM is unique in its ability to provide closed-form solutions that exactly satisfy Maxwell's equations. As a result, the solutions can be Lorentz transformed without introducing additional error. 

In this article, we derive exact solutions to Maxwell's equations for the electromagnetic fields of a constant-velocity flying focus pulse. The approach combines the CSPM with a Lorentz transformation from a frame in which the focus is stationary to a frame in which the focus is moving. The vector solutions are inherently non-paraxial, can have arbitrary polarization, and are generalized to higher-order radial and orbital angular momentum modes. Subluminal ($|\text{v}_I| < c$) and superluminal ($|\text{v}_I| > c$) solutions are constructed from multipole spherical and hyperbolic wave solutions, respectively. Propagating the fields backward in space reveals that each solution corresponds to a pulse that was focused by a lens with a time-dependent focal length. Thus, these solutions can be experimentally realized using the flying focus X. For a wide range of parameters, the differences between the exact solutions and simpler paraxial solutions are small, justifying the use of paraxial solutions for theoretical or computational studies of flying focus applications in many regimes. These solutions also compliment the larger body of work on arbitrary velocity autofocusing and nondiffracting waves, e.g. Refs. \cite{saari2020relativistic,besieris2022autofocusing,Yessenov22}.

The remainder of this article is organized as follows. In Sec. \ref{sec:solutions}, multipole solutions to Maxwell's equations are derived and then transformed into beam-like solutions using the CSPM. Guided by the insights of Refs. \cite{longhi2004gaussian} and \cite{belanger1986lorentz}, the exact fields of subluminal and superluminal foci are found by Lorentz transforming the beam-like solutions. Section \ref{sec:ExpSols} presents explicit modal solutions for the four-potential of moving foci, including an example with orbital angular momentum. Section \ref{sec:prop&foc} describes the procedure for constructing pulsed solutions from a superposition of modal solutions and determines the optical assembly required to produce the pulsed fields in an experiment. Section \ref{sec:PXvsHZ} compares the exact solutions to paraxial solutions and provides conditions for the reliable use of paraxial solutions when modeling a flying focus. Section \ref{sec:Conclusions} summarizes the results.

\begin{figure*}
\centering\includegraphics[width=5in]{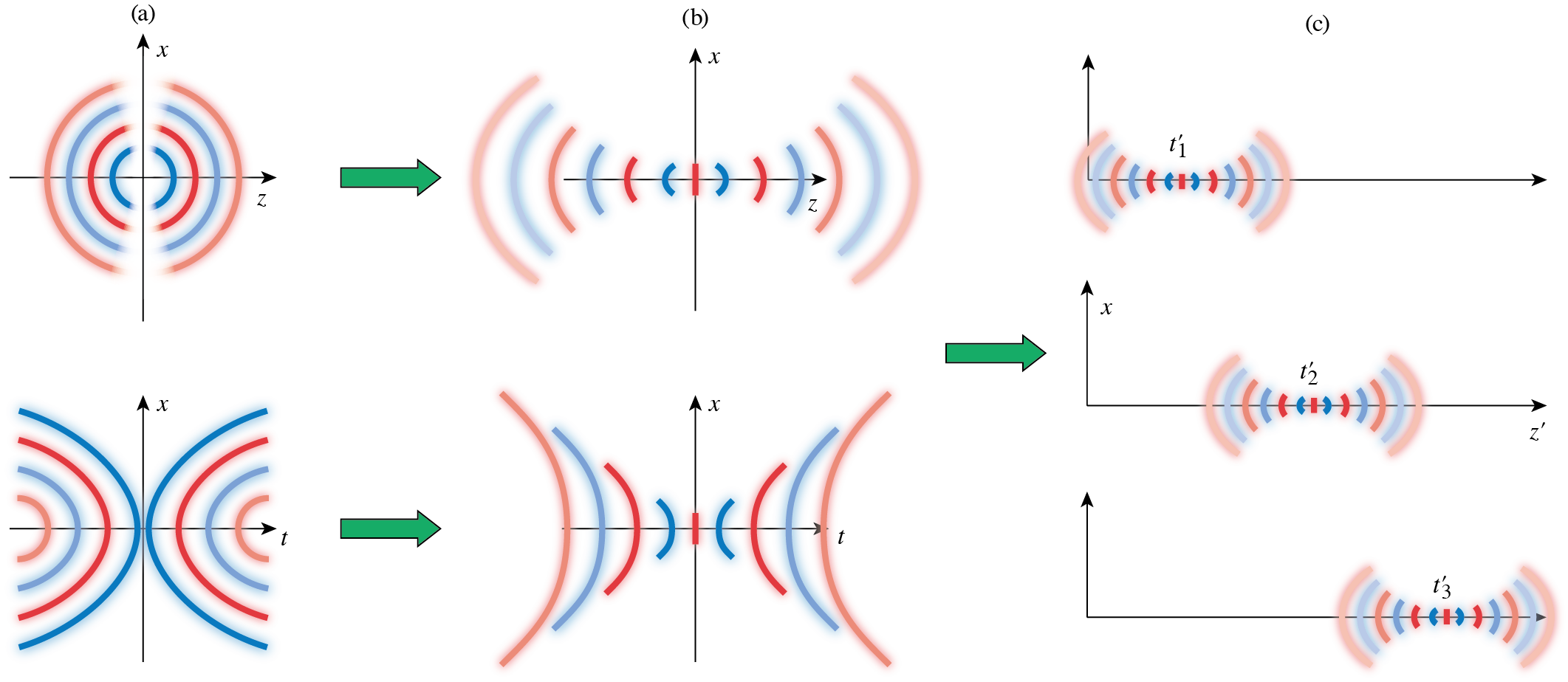}
\caption{A schematic of the theoretical approach. (a) The approach starts with a multipole spherical (top) or hyperbolic (bottom) solution. (b) A displacement of the coordinate $z$ (top) or $t$ (bottom) into its complex plane transforms the multipole solution into a beam-like solution with a stationary focus in space (top) or time (bottom). (c) A Lorentz transformation of either beam-like solution from a frame of reference in which the foci appear stationary to the laboratory frame produces the exact electromagnetic fields of a flying focus.}
\label{fig:1}
\end{figure*}

\section{Lorentz transformations of complex source point fields}\label{sec:solutions}

In vacuum, the electromagnetic fields, the four-potential in the Lorenz gauge, and the Hertz vectors, all satisfy the homogeneous wave equation. Consider a scalar solution $u(\bvec{x},t)$ to the wave equation,
\begin{equation}
    \left( \nabla_\perp^2+\partialDsq{z} - \partialDsq{t} \right)u(\bvec{x},t) = 0,
\end{equation}
where the speed of light $c=1$. The solution can be written as a superposition of modal solutions with explicit harmonic dependence in either time or space: $u(\bvec{x},t) = \int{u_{\kappa}(\bvec{x},t)} \, d \kappa$, where $u_{\kappa}(\bvec{x},t) = \frac{1}{2}S(\bvec{x}_\perp,z)e^{-i\kappa t} + \mathrm{c.c.}$ or $u_{\kappa}(\bvec{x},t) = \frac{1}{2}H(\bvec{x}_\perp,t)e^{i\kappa z} + \mathrm{c.c.}$. Substituting these into the wave equation yields
\begin{subequations}\label{eq:HelmLikeEqs} 
\begin{align} 
\left( \nabla^2_\perp+\partialDsq{z}+ \kappa^2 \right)S(\bvec{x}_\perp,z) &=0,   \label{eq:Helms}\\ 
\left(\nabla^2_\perp - \partialDsq{t}- \kappa^2\right) H(\bvec{x}_\perp,t) &= 0.\label{eq:HyperHelm} 
\end{align} 
\end{subequations}
Equations \eqref{eq:Helms} and \eqref{eq:HyperHelm} are the Helmholtz equation and its hyperbolic analog, respectively. The solutions are multipole spherical ($S$) and hyperbolic ($H$) waves
\begin{subequations}\label{eq:Sols}
\begin{align} 
S(\bvec{x}_\perp,z) = \sum_{n}{\sum_{\ell=-n}^{n} {\alpha_{n \ell}(\kappa ) j_n(\kappa R) P_n^\ell(\cos \phi)e^{i\ell\theta} }},   \label{eq:SgenSol}\\ 
H(\bvec{x}_\perp,t) = \sum_{n}{\sum_{\ell=-n}^{n}{ \alpha_{n \ell}(\kappa)k_n(\kappa \mathscr{R} \mathscr{a}) P_n^\ell(\cos \varphi)e^{i\ell\theta}}},\label{eq:HgenSol} 
\end{align} 
\end{subequations}
where $\alpha_{n \ell}(\kappa )$ is a weighting factor, $j_n$ is the $n^\text{th}$ spherical Bessel function of the first kind, $k_n$ is the modified spherical Bessel function of the second kind, $P_n^\ell$ is the associated Legendre polynomial, $\ell$ is the azimuthal mode number, $R= (\rho^2+z^2)^{1/2}$, $\mathscr{R}=(\rho^2-t^2)^{1/2}$, $\rho = |\bvec{x}_\perp|$, $\cos(\phi) = z/R$,  $\cos(\varphi) =it/\mathscr{R}$, and $\theta = \arctan (y/x)$. Spherical Bessel functions of the second kind have been omitted in Eq. \eqref{eq:SgenSol} because they result in real-valued branch points when using the CSPM \cite{heyman2001gaussian,sheppard1999electromagnetic}. Modified spherical Bessel functions of the first kind have been omitted in Eq. \eqref{eq:HgenSol} because they diverge as $\mathscr{R} \rightarrow \infty$. 

Hertz vectors provide a convenient mathematical representation for calculating the four-potential or electromagnetic fields. With a closed-form expression for a single vector component of the Hertz vectors, one can generate the entire four-potential and all six components of the electromagnetic field by taking derivatives. In particular, a multipole spherical or hyperbolic wave that propagates outward from the origin [Fig. \ref{fig:1}(a)] can be formed by using equal and crossed electric and magnetic Hertz vectors: 
\begin{equation}\label{eq:HZvecs}
    \begin{split}
    \bm{\Pi}_e(\bvec{x},t) &=  u_{\kappa}(\bvec{x},t)\unitvec{e}, \\
    \bm{\Pi}_m(\bvec{x},t) &=  u_{\kappa}(\bvec{x},t)\unitvec{m}, \\
    \end{split}
\end{equation}
where $\unitvec{e}\cdot\unitvec{m}=0$ and $\unitvec{e}\times\unitvec{m}=\unitvec{z}$ \cite{cullen1979complex}. With this configuration, $\unitvec{z}$ and $\unitvec{e}$ determine the predominate directions of propagation and electric-field polarization, respectively. 

A spherical or hyperbolic solution $u_\kappa(\bvec{x},t)$ remains a solution to the homogeneous wave equation under a coordinate translation along the real \emph{or imaginary} axis. Displacing a coordinate into its complex plane transforms a multipole spherical or hyperbolic wave into a beam-like wave, in which the phase fronts pass through the origin instead of originating from it [Fig. \ref{fig:1}(b)] \cite{deschamps1971gaussian,shin1977gaussian, zauderer1986complex, norris1986complex,heyman2001gaussian,cullen1979complex,sheppard1999electromagnetic,mitri2013quasi}. This is the CSPM. For the spherical solutions, a beam-like wave is generated by transforming the axial coordinate as $z \rightarrow z - iZ_R$, such that $S(\bvec{x}_\perp,z) \rightarrow S(\bvec{x}_\perp,z-iZ_R)$. For hyperbolic solutions, a beam-like wave is generated by transforming time as $t \rightarrow t - iZ_R$, such that $H(\bvec{x}_\perp,t) \rightarrow H(\bvec{x}_\perp,t-iZ_R)$. In the paraxial limit, i.e., when $|z- iZ_R|$ or $|t- iZ_R|\gg \rho$, the minimum spot size of the beam-like wave is given by $w_0 = (2Z_R/\kappa)^{1/2}$, thus $Z_R = \tfrac{1}{2}\kappa w_0^2$ corresponds to the Rayleigh range. When working with beam-like waves, it is convenient to introduce the complex beam parameter $q$. In the context of spherical and hyperbolic solutions, $q(z) = z-\tfrac{1}{2}i\kappa w_0^2$ and $q(t) = t-\tfrac{1}{2}i\kappa w_0^2$, respectively. 

The exact solutions provided by the CSPM, i.e.,
\begin{equation} \label{eq:CSPMexact} 
    \begin{split}
    \bm{\Pi}_e(\bvec{x},t) &=  \tfrac{1}{2}S\bigl(\bvec{x}_\perp,q(z)\bigr)e^{-i\kappa t} \unitvec{e} \, + \, \mathrm{c.c.} \\
    \bm{\Pi}_e(\bvec{x},t) &=  \tfrac{1}{2}H\bigl(\bvec{x}_\perp,q(t)\bigr)e^{i\kappa z} \unitvec{e} \, + \, \mathrm{c.c.},  \\
    \end{split}
\end{equation}
with $\bm{\Pi}_m = \unitvec{z} \times \bm{\Pi}_e$, describe continuous-wave laser fields with stationary foci [Fig. \ref{fig:1}(b)]. Electromagnetic fields that satisfy Maxwell's equations in one frame of reference satisfy Maxwell's equations in all other inertial reference frames. Therefore, there exists a frame of reference in which the focus appears to be moving at a velocity $\text{v}_I = c\beta_I$. In the context of a flying focus, this frame with a moving focus is the laboratory frame. 

When performing a Lorentz transformation from the stationary frame to the laboratory frame, it is convenient to work with the four-potential $A^{\mu} = (\Phi,\bm{\mathrm{A}})$. The four-potential in the stationary frame can be calculated from the Hertz vectors in Eq. \eqref{eq:CSPMexact} as follows \cite{jackson_classical_1999}:
\begin{equation}\label{eq:4vecCalc}
   \begin{split}
    \Phi(\bvec{x},t) &= -\nabla \cdot \bm{\Pi}_e(\bvec{x},t),\\
   \bm{\mathrm{A}}(\bvec{x},t) &= \partialD{t} \bm{\Pi}_e(\bvec{x},t) +\nabla \times \bm{\Pi}_m(\bvec{x},t).\\
   \end{split}
\end{equation}
Because the Hertz vectors are formulated in the Lorenz gauge \cite{Nisbet1955Hertz}, the condition $\nabla \cdot \bm{\mathrm{A}} +\partial_t \Phi = 0$ is automatically satisfied. Further, the relationship $\bm{\Pi}_m= \unitvec{z}\times\bm{\Pi}_e$ implies that $A_z = -\Phi$, resulting in the Lorentz invariant four-vector dot product $A^{\mu}A_{\mu} = -\Phi^2+A_\perp^2 + A_z^2 = A_\perp^2$. The four-potential in the laboratory frame (denoted by a prime $'$) is given by
\begin{equation}\label{eq:PhiPrime}
   \begin{split}
    \Phi'(\bvec{x}',t') &= \gamma (1 - \beta)\Phi(\bvec{x},t),
   \end{split}
\end{equation}
$\textbf{A}'_{\perp} = \textbf{A}_{\perp}$, and $A'_z = -\Phi'$, where $\gamma = (1-\beta^2)^{-1/2}$ is the Lorentz factor. Note that in the laboratory frame, the stationary frame appears to be moving at a velocity $-\beta$. The definition of this velocity and the mapping between  ($\bm{\mathrm{x}},t)$ and ($\bm{\mathrm{x}}',t')$ depend on whether the focal velocity is subluminal $|\beta_I| < 1$ or superluminal $|\beta_I| > 1$.

Upon Lorentz transforming to the laboratory frame, the spherical solutions describe foci that move at subluminal velocities $|\beta_I| < 1$. In this case, $\beta = \beta_I$ and the coordinates transform as
\begin{equation}\label{eq:Scoords}
    \begin{split}
    t &= \gamma_{\scaleto{<}{3pt}} (t'-\beta_I z'),\\
    q(z) &= \gamma_{\scaleto{<}{3pt}} (z'-\beta_I t')-\frac{i\kappa' w_0^2}{2\gamma_{\scaleto{<}{3pt}}(1+\beta_I)},
    \end{split}
\end{equation}
where $\gamma_{\scaleto{<}{3pt}} = (1-\beta_I^2)^{-1/2}$ and $\kappa' = \gamma_{\scaleto{<}{3pt}}(1+\beta_I)\kappa$ is the laboratory frame value of $\kappa$. The hyperbolic solutions describe foci that move at superluminal velocities $|\beta_I| > 1$. Clearly a Lorentz transformation using a $|\beta| > 1$ would be unphysical. Nevertheless, a superluminal focus can be achieved by Lorentz transforming the hyperbolic solutions using $\beta = \beta_I^{-1}$, such that
\begin{equation}\label{eq:Hcoords}
    \begin{split}
    q(t) &= \gamma_{\scaleto{>}{3pt}} (t' - z'/\beta_I)-\frac{i\beta_I\kappa' w_0^2}{2\gamma_{\scaleto{>}{3pt}}(1+\beta_I)}\\
    z &= \gamma_{\scaleto{>}{3pt}} (z' - t'/\beta_I),
    \end{split}
\end{equation}
where, in this case, $\gamma_{\scaleto{>}{3pt}} = (1-\beta_I^{-2})^{-1/2}$ and $\kappa' = \gamma_{\scaleto{>}{3pt}}(1+\beta_I^{-1})\kappa$.

For both spherical and hyperbolic waves, the focal plane, defined by $\mathrm{Re}(q)=0$, travels along the trajectory $z'=\beta_I t'$ [Fig. \ref{fig:1}(c)]. The time that it takes the confocal region to move past a fixed point in space, i.e., the duration of the moving focus $\tau$, is obtained from the time scale evident in the expressions for $q$. Specifically setting $|q|=z'=0$ and solving for $t'$, one finds
\begin{equation}\label{eq:duration}
    \tau = \Bigl|\frac{1-\beta_I}{\beta_I}\Bigr| Z'_R,
\end{equation}
where $Z'_R = \tfrac{1}{2}\kappa' w_0^2$ is the Rayleigh range in the laboratory frame and $\beta_I$ can take any value other than 1. The duration is identical to that of an intensity peak produced by a lens with a focal length that depends linearly on time, as in the flying focus X \cite{simpson2022spatiotemporal}.

Once the potentials have been calculated using Eqs. \eqref{eq:CSPMexact} -- \eqref{eq:Hcoords}, the electromagnetic fields can be found in the usual way: $\bm{\mathrm{E}}' = -\nabla'\Phi' - \partial_{t'} \bm{\mathrm{A}}' $ and $\bm{\mathrm{B}}' = \nabla' \times \bm{\mathrm{A}}' $. As will be shown below, all six components of the resulting fields have  nonzero values. Further, the transverse components of the electric and magnetic field that are perpendicular to \unitvec{e} and \unitvec{m}, respectively, are equal---a symmetry which is frequently sought-after in solutions to Maxwell's equations \cite{levy2019mathematics,barton1989fifth,quesnel1998theory,sheppard1999electromagnetic}. The relative amplitudes of the field components scale as $|\unitvec{e} \cdot \bvec{E}'| \sim |\unitvec{m} \cdot \bvec{B}'|$ and
\begin{equation}\label{eq:AmpScale}
    |\unitvec{z} \cdot \bvec{E}'| \sim |\unitvec{z} \cdot \bvec{B}'| \sim \mathrm{min} \left( \frac{1}{\kappa'w_0} , \, |1-\beta_I|\kappa'w_0 \right) |\unitvec{e} \cdot \bvec{E}'|.
\end{equation}
For circular polarization, these two scalings are sufficient. Linear polarization has the additional scaling
\begin{equation}\label{eq:AmpScale2}
    \begin{split}
    |\unitvec{m} \cdot \bvec{E}'| \sim |\unitvec{e} \cdot \bvec{B}'| \sim \mathrm{min} \left( \frac{1}{\kappa'w_0} , \, |1-\beta_I|\kappa'w_0 \right) |\unitvec{z} \cdot \bvec{E}'|.
    \end{split}
\end{equation}
When $\beta_I = 0$, these scalings reduce to those of a stationary focus. In the limit as $\beta_I \rightarrow 1$, the duration $\tau$ becomes shorter than $2\pi/ \kappa'$, and the components of the electric and magnetic fields orthogonal to $\unitvec{e}$ and $\unitvec{m}$ vanish.

\section{Explicit solutions}\label{sec:ExpSols}

This section presents expressions for the four-potential of arbitrary-velocity subluminal and superluminal foci. The expressions exactly satisfy the wave equation. In each example, the four-potential is derived from a single modal solution $u_{\kappa}$. Explicit expressions for each component of the electromagnetic field can be found directly from the four-potential, but are unwieldy and provide little additional insight. Instead, the structure of the electromagnetic fields is highlighted by figures.

\begin{figure*}
\centering\includegraphics[width=6.75in]{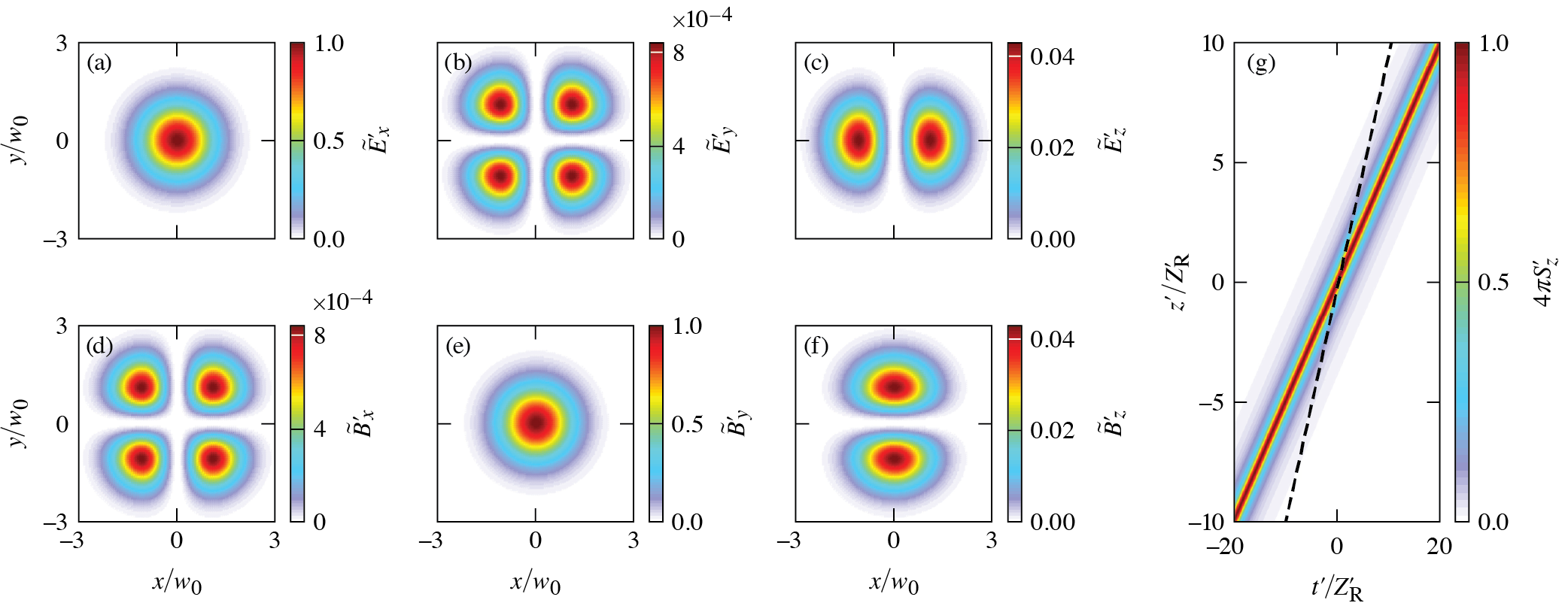}
\caption{A subluminal focus with $\beta_I = 0.5$, $n = \ell = 0$, and $\kappa' w_0 = 20$. [(a)--(f)] Cross sections of the electromagnetic field amplitudes at the location of the moving focus $z'=\beta_I t'$. Here $\Tilde{E'} = \langle E^{\prime 2} \rangle ^ {1/2} $, where $\langle \rangle$ denotes a cycle average. 
The amplitudes are normalized to the $\mathrm{max}(\Tilde{E_x'})$. (g) The longitudinal component of the cycle-averaged Poynting vector $S_z'$ at $\rho = 0$, showing the motion of the focus. The dashed black line demarcates the trajectory $z'=t'$ for reference.}
\label{fig:2}
\end{figure*}

\subsection{Subluminal focus}\label{subsec:Ex1}
Expressions for the four-potential of subluminal foci are obtained from the spherical solutions. As a first example, consider the lowest order radial and azimuthal mode ($n = \ell = 0$) in Eq. \eqref{eq:SgenSol}. Upon choosing the polarization vector $\unitvec{e} = \unitvec{x}$, the electric and magnetic Hertz vectors are given by
\begin{equation} \label{eq:example1}
    \bm{\Pi}_e(\bvec{x},t)=\tfrac{1}{2}\alpha_{00} j_0(\kappa R)e^{-i\kappa t}\unitvec{x} +\text{c.c.}\\
\end{equation}
and $\bvec{\Pi}_m = \unitvec{z}\times \bm{\Pi}_e$, respectively. After applying the CSPM, Eq. \eqref{eq:4vecCalc}, and a Lorentz transformation to Eq. \eqref{eq:example1}, one finds the laboratory frame four-potential
\begin{equation} \label{eq:sub00sol}
    \begin{split}
    \Phi' &=  \frac{\alpha_{00}(1-\beta_I)\gamma_{\scaleto{<}{3pt}}}{2}\frac{\kappa x}{R}j_1(\kappa R)e^{-i\kappa t} +\text{c.c.},\\[10pt]
    A_x' &=-\frac{i\alpha_{00}}{2} \kappa  \left[j_0(\kappa R)+i\frac{q(z)}{R} j_1(\kappa R) \right]e^{-i\kappa t}+\text{c.c.},
    \end{split}
\end{equation}
$A_y' = 0$, and $A_z' = -\Phi'$, where $\kappa = \gamma_{\scaleto{<}{3pt}}(1-\beta_I)\kappa'$ and $R = \sqrt{\rho^2 + q^2(z)}$ with $q(z)$ and $t$ given by Eq. \eqref{eq:Scoords}. For a stationary focus with $\beta_I = 0$, the fields derived from Eq. \eqref{eq:sub00sol} are identical to those in Ref. \cite{sheppard1999electromagnetic}.

Figures \ref{fig:2}(a) -\ref{fig:2}(f) display cross sections of the resulting electric and magnetic fields at the location of the moving focus $z'=\beta_I t'$ for $\beta_I = 0.5$ and $\kappa'w_0 = 20$. The predominant electric and magnetic fields, $E'_x$ and $B'_y$, have equal amplitudes and Gaussian-like transverse profiles. The remaining vector components exhibit more complex spatial structure, but are much smaller in amplitude, consistent with Eqs. \eqref{eq:AmpScale} and \eqref{eq:AmpScale2}. Figure \ref{fig:2}(g) illustrates the motion of the focus in the laboratory frame. The cycle-averaged longitudinal component of the Poynting vector, $S_z' = \unitvec{z}\cdot \langle\bvec{E}'\times\bvec{B}'\rangle/4\pi$, is plotted as a function of $z'$ and $t'$ at $\rho = 0$. As expected from Eqs. \eqref{eq:Scoords} and \eqref{eq:duration}, the peak of $S_z'$ travels at the velocity $\beta_I$ and has a duration $\tau = Z_R'$. For comparison, the dashed black line demarcates the speed of light trajectory $z' = t'$.

\begin{figure*}
\centering\includegraphics[width=6.75in]{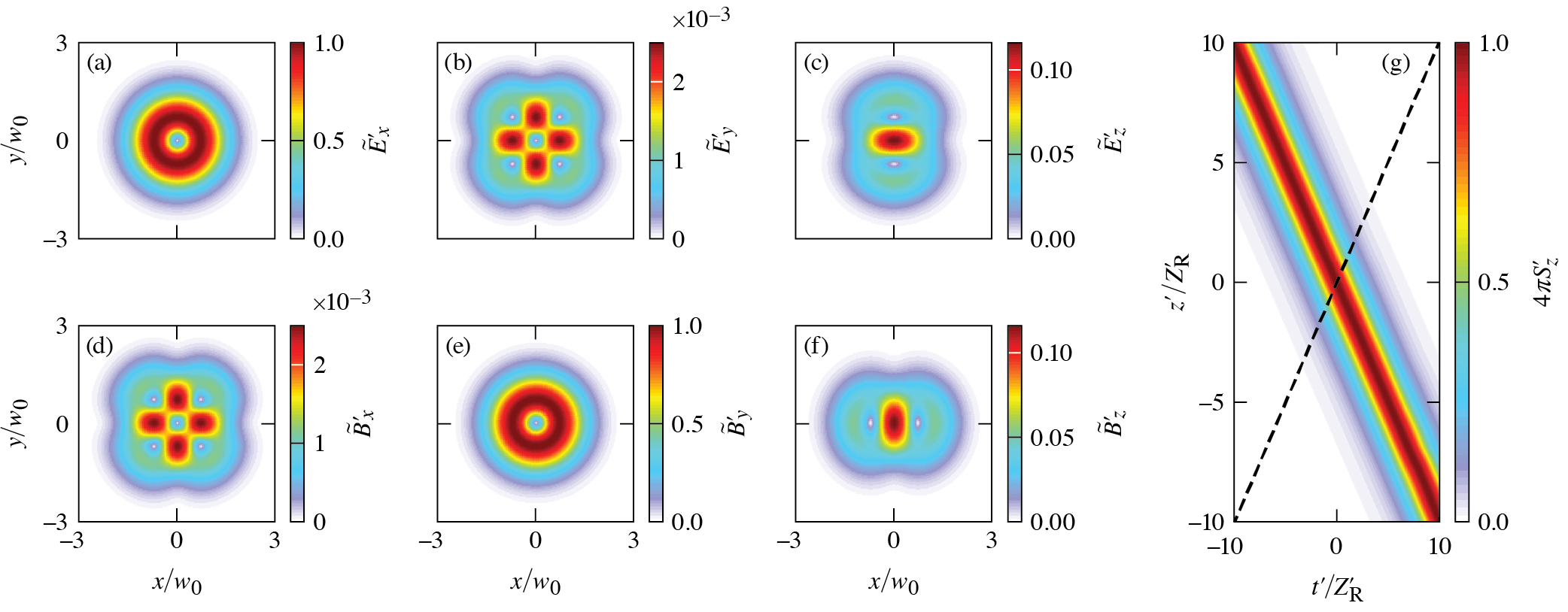}
\caption{A backward focus carrying orbital angular momentum with $\beta_I = -0.99$, $n = \ell = 1$, and $\kappa' w_0 = 20$. [(a)--(f)] Cross sections of the electromagnetic field amplitudes at the location of the moving focus $z'=\beta_I t'$. Here $\Tilde{E'} = \langle E^{\prime 2} \rangle ^ {1/2} $, where $\langle \rangle$ denotes a cycle average. The amplitudes are normalized to the $\mathrm{max}(\Tilde{E_x'})$.  (g) The longitudinal component of the cycle-averaged Poynting vector $S_z'$ at $\rho = 0$, showing the motion of the focus. The dashed black line demarcates the trajectory $z'=t'$ for reference.}
\label{fig:3}
\end{figure*}

A moving focus carrying orbital angular momentum can be described by any solution with $|\ell|>0$. The spherical solution with $\ell = n = 1$ will be used as an example. The electric and magnetic Hertz vectors for this mode are given by 
\begin{equation} \label{eq:example2}
    \bm{\Pi}_e(\bvec{x},t)=-\tfrac{1}{2}\alpha_{11} j_1(\kappa R)\sin(\phi) e^{i\ell\theta-i\kappa t}\unitvec{x} +\text{c.c.}
\end{equation}
and $\bvec{\Pi}_m = \unitvec{z}\times \bm{\Pi}_e$, where $\unitvec{e} = \unitvec{x}$ has been chosen for the polarization vector. Following the same procedure described above provides
\begin{equation} \label{eq:sub11sol}
    \begin{split}
    \Phi' &=  \frac{ \alpha_{11}(1-\beta_I)\gamma_{\scaleto{<}{3pt}}}{ 2R } \Bigl[  
     j_1(\kappa R) - \frac{\kappa \rho^2}{R} \cos(\theta) e^{i\theta} j_2(\kappa R) \Bigr]e^{-i \kappa t} \\ &+\mathrm{c.c.},\\[8pt]
    A_x' &= \frac{i\alpha_{11}}{2}\frac{\kappa \rho}{R} \left[j_1( \kappa R) + i\frac{q(z)}{R}  j_2( \kappa R) \right] e^{i\theta-i \kappa t} +\text{c.c.},
    \end{split}
\end{equation}
$A_y' = 0$, and $A_z' = -\Phi'$. Figure \ref{fig:3} shows cross sections of the fields and the longitudinal Poynting vector for the case of a backward focus with $\beta_I = -0.99$ and $\kappa' w_0 = 20$. The predominant field components, $E'_x$ and $B'_y$, exhibit the characteristic donut-like profile with a maximum amplitude located at $\rho \approx w_0/\sqrt{2}$. As before, the other field components exhibit more structure, but are much smaller in amplitude. In contrast to the $\ell = 0$ example, the longitudinal field components of the moving focus are nonzero at $\rho =0$. Note that the longitudinal Poynting vector is positive despite the backward motion of the peak.

\subsection{Superluminal focus}\label{subsec:Ex2}
\begin{figure*}
\centering\includegraphics[width=6.75in]{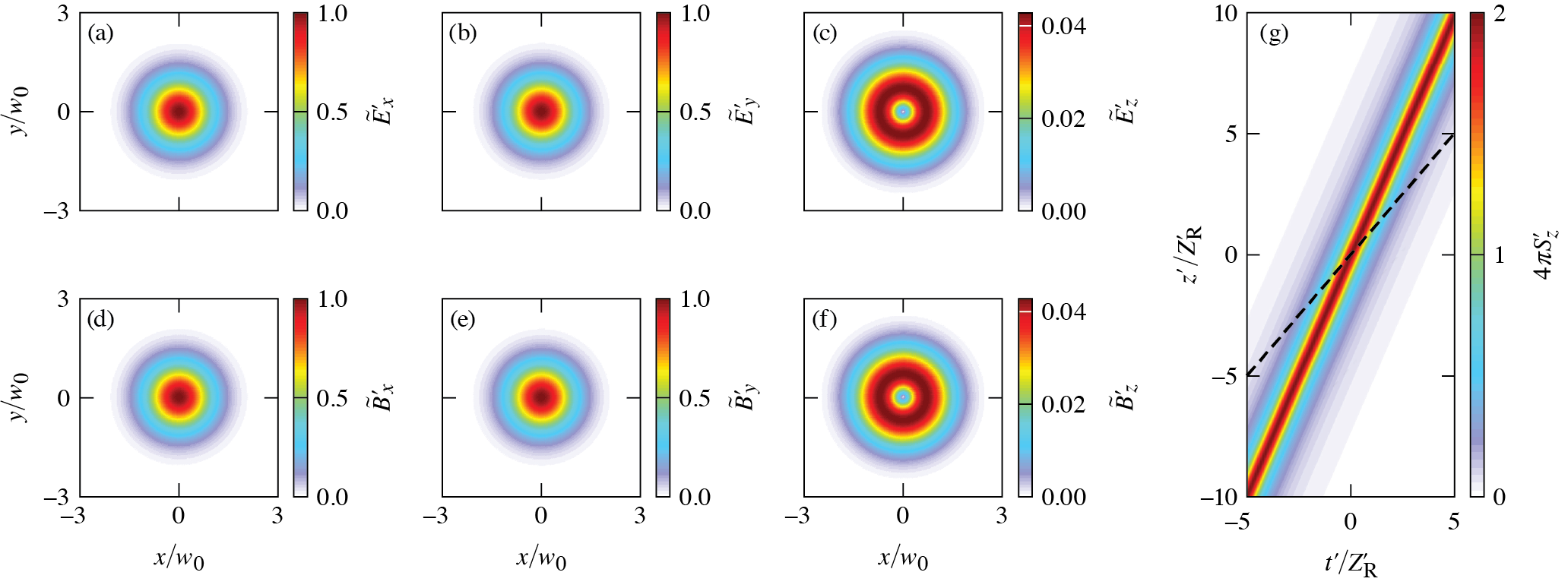}
\caption{A circularly polarized superluminal focus with $\beta_I = 2$, $n = \ell = 0$, and $\kappa' w_0 = 20$. [(a)--(f)] Cross sections of the electromagnetic field amplitudes at the location of the moving focus $z'=\beta_I t'$. Here $\Tilde{E'} = \langle E^{\prime 2} \rangle ^ {1/2} $, where $\langle \rangle$ denotes a cycle average. The amplitudes are normalized to the $\mathrm{max}(\Tilde{E_x'})$. (g) The longitudinal component of the cycle-averaged Poynting vector $S_z'$ at $\rho = 0$, showing the motion of the focus. The dashed black line demarcates the trajectory $z'=t'$ for reference.}
\label{fig:4}
\end{figure*}

Expressions for the four-potential of superluminal foci are obtained from the hyperbolic solutions. For the simplest example, consider the lowest order radial and azimuthal mode ($n = \ell = 0$) of the hyperbolic solution. The electric and magnetic Hertz vectors are
\begin{equation}  
    \bm{\Pi}_e(\bvec{x},t)=\tfrac{\sqrt{2}}{4}\alpha_{00}k_0(\kappa \mathscr{R})e^{i\kappa z}(\unitvec{x}+i\unitvec{y})+\text{c.c.},\\
\end{equation}
and $\unitvec{z}\times\bm{\Pi}_e$, respectively. Here circular polarization, i.e., $\unitvec{e}=\tfrac{1}{\sqrt{2}}(\unitvec{x}+i\unitvec{y})$, has been chosen to demonstrate the generality of the solutions to describe polarizations other than linear. Upon using the CSPM, Eq. \eqref{eq:4vecCalc}, and a Lorentz transformation, one finds the laboratory frame four-potential
\begin{equation}\label{eq:sup00sol}
    \begin{split}
    \Phi' &=    \frac{\sqrt{2}\alpha_{00}(1-1/\beta_I) \gamma_{\scaleto{>}{3pt}}}{4}
    \frac{\rho}{\mathscr{R}^2} (1 + \kappa\mathscr{R} ) 
 k_0(\kappa\mathscr{R})e^{i\theta+i\kappa z}+\text{c.c.}, \\[10 pt]
    A_x' &= -\frac{i\sqrt{2}\alpha_{00}}{4}\kappa\left[1 + i\frac{q(t)}{\mathscr{R}} + i \frac{q(t)}{\kappa \mathscr{R}^2} \right]k_0(\kappa\mathscr{R})e^{i\kappa z}+\text{c.c.},\\
    \end{split}
\end{equation}
$A_y' = iA_x'$, and $A_z' = -\Phi'$, where $\kappa = \gamma_{\scaleto{>}{3pt}}(1-1/\beta_I)\kappa'$ and $\mathscr{R} = \sqrt{\rho^2 - q^2(t)}$ with $q(t)$ and $t$ given by Eq. \eqref{eq:Hcoords}. Figure \ref{fig:4} displays cross sections of the resulting fields and the longitudinal Poynting vector for $\beta_I = 2$ and $\kappa' w_0 = 20$. The use of circular polarization results in identical, near-Gaussian profiles for each of the transverse field components and symmetric donut-like profiles for the longitudinal components. The peak of the longitudinal Poynting vector follows the trajectory $z' = \beta_I t'$ and has a duration $\tau$. 

\subsection{Luminal focus}\label{subsec:Ex3}

Expressions for the four-potential of luminal foci can be found by taking the limit of the subluminal and superluminal solutions as $|\beta_I| \rightarrow 1$. The limit as $|\beta_I| \rightarrow 1$ is identical from above and below, and the sub- and superluminal solutions reduce to the same expression. When $\beta_I = -1$, the transverse vector potential exactly satisfies a paraxial wave equation, and thus can be expressed as a  Laguerre--Gaussian mode (see Sec. \ref{sec:PXvsHZ}). Taking the limit of the linearly polarized $n=\ell=0$ sub or superluminal solution as $\beta_I \rightarrow 1$ yields
\begin{equation}\label{eq:fromBelow}
    A_x' \propto j_1\bigl(\kappa'(z'-t')\bigr)
\end{equation}
and $\Phi' = A_z'  = A_y' = 0$. Equation \eqref{eq:fromBelow} is independent of the transverse coordinates, i.e., it is a plane wave with $E_x' = B_y'$ and all other field components equal to zero [see Eqs. \eqref{eq:AmpScale} and \eqref{eq:AmpScale2}]. 

As $\beta_I \rightarrow1$, the complex coordinate introduced in the CSPM approaches the real axis. The absence of the imaginary term in Eqs. \eqref{eq:Scoords} and \eqref{eq:Hcoords} eliminates the beam-like behavior of the solutions and results in the plane wave solution [Eq. \eqref{eq:fromBelow}]. With a plane wave, there is no distinction between the near and far fields. Even when taken in superposition, the laser pulse and focal plane would have to coincide everywhere in space for all time, rendering the plane wave solution impossible to produce in this configuration. Nevertheless, a flying focus with $\beta_I=1$ can be generated using other techniques like the achromatic flying focus or arbitrary structured laser (ASTRL) pulses \cite{palastro2020dephasingless,pierce2022ASTRL}.

\section{Pulsed solutions and focal range}\label{sec:prop&foc}
In every example presented in Sec. \ref{sec:ExpSols}, the electromagnetic fields were generated from a single modal solution $u_{\kappa}$. The modal solutions used to generate subluminal foci are localized in space, but oscillate at a single period $2\pi/\kappa$ for all time; The modal solutions used to generate the superluminal foci are localized in time, but oscillate at a single wavelength $2\pi/\kappa$ everywhere on the $z$ axis. In both cases, the focus travels along the trajectory $z'=\beta_It'$ forever, and the electromagnetic energy is infinite. Physically realizable electromagnetic fields are localized in space and time and have finite energy, i.e., they are pulsed. Such fields can be generated from a discrete or continuous superposition of the modal solutions, $u(\bvec{x},t) = \int{u_{\kappa}(\bvec{x},t)} \, d\kappa$. 

The temporal or longitudinal profile of a pulse depends on the spectral amplitude and phase of each modal solution in the superposition. The spectral amplitudes and phases, i.e., the $\alpha_{n\ell}(\kappa)$, can be chosen such that a single component of either the Hertz vectors, four-potential, or fields exhibits a particular temporal or longitudinal profile. For consistency with the previous section, the $\alpha_{n\ell}(\kappa)$ will be chosen to specify the profile of the predominant component of the four-potential in the laboratory frame $A'_x$. 

To begin, note that an $A'_x$ derived from a single modal solution always takes the form of either $A'_x = \alpha_{n\ell}(\kappa)f_{n\ell}(\bvec{x}_{\perp},z;\kappa)e^{-i\kappa t} + \text{c.c.}$ for subluminal foci or $A'_x = \alpha_{n\ell}(\kappa)f_{n\ell}(\bvec{x}_{\perp},t;\kappa)e^{i\kappa z}  + \text{c.c.}$ for superluminal foci. The space--time location of the maximum of $|f_{n\ell}|$ is insensitive to $\kappa$ (and is fully independent of $\kappa$ in the paraxial limit). As a result, the choice 
\begin{equation}
\alpha_{n\ell}(\kappa) = \frac{a_0 g_{n\ell}(\kappa)}{ \mathrm{max}|f_{n\ell}|} 
\end{equation}
ensures that a pulse with a subluminal or superluminal focus has a temporal profile $\hat{g}_{n\ell}(t)=\int{g_{n\ell}(\kappa)e^{-i\kappa t}} \, d\kappa$ or longitudinal profile $\hat{g}_{n\ell}(z)=\int{g_{n\ell}(\kappa)e^{i\kappa z}} \, d\kappa$, respectively. 
Further, if $|\hat{g}_{n\ell}(0)|=1$, then the transverse vector potential in the laboratory frame will have a maximum amplitude of $a_0$. To ensure that the moving focus has a near-constant profile and maximum Poynting vector over its entire trajectory, $\hat{g}_{n\ell}$ should have a near-flattop profile. In this work, $\hat{g}_{n\ell}(t) = \exp{[-(t/T)^4-i\kappa_0t]}$ and $\hat{g}_{n\ell}(z) = \exp{[-(z/T)^4+i\kappa_0z]}$ are used. 

The pulsed fields generated by a superposition of modal solutions exhibit a moving focus over a finite duration and spatial extent. The length $L$ over which the focus persists, i.e., the focal range, is determined by the transit time $\Delta T$ of the focus through the entire pulse duration $T$. In vacuum, the pulse propagates at its group velocity $c=1$, such that
$\Delta T = T/|1-\beta_I|$. Over the interval $\Delta T$, the focus travels a distance $L =  |\beta_I|\Delta T$, providing
\begin{equation} \label{eq:Lexp}
    L = \Big |\frac{\beta_I}{1-\beta_I} \Big | T.
\end{equation}
Equation \eqref{eq:Lexp} demonstrates that the focal range $L$ increases with $T$, approaches $T$ as $|\beta_I| \rightarrow \infty$, and is much greater than $T$ when $\beta_I \approx 1$. In addition, Eq. \eqref{eq:Lexp} can be combined with Eq. \eqref{eq:duration} to show that
$\tau/T = Z'_R/L$. Note that every wavenumber or frequency $\kappa'$ within the superposition will have a different Rayleigh length, $\tfrac{1}{2}\kappa' w_0^2$. For this section, the quantity $Z'_R$ is defined in terms of the central frequency or wavenumber $\kappa'_0$, such that $Z'_R = \tfrac{1}{2}\kappa'_0 w_0^2$.

\begin{figure}[H]
\centering\includegraphics[width=3.375in]{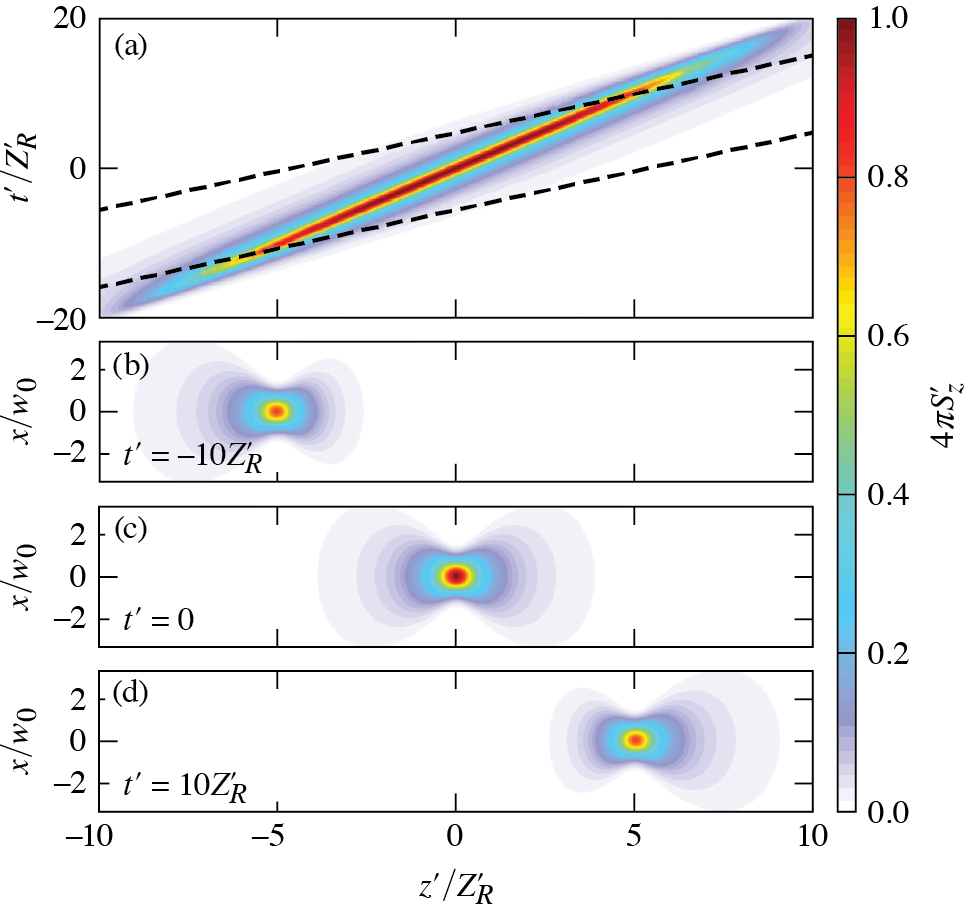}
\caption{A pulsed subluminal focus constructed from a superposition of modal solutions with $\beta_I = 0.5$, $n = \ell = 0$, $\kappa'_0 w_0 = 20$, and $T=10Z'_{R}$ ($L = 10Z'_{R}$). (a) The longitudinal component of the cycle-averaged Poynting vector $S_z'$ at $\rho = 0$, showing the motion and finite extent of the focus [cf., Fig. \ref{fig:2}(g)]. The dashed black lines mark the full width at half maximum of the pulse $\hat{g}_{00}$, which travels at the speed of light. [(b)--(d)] The profile of the longitudinal Poynting vector in the $x-z'$ plane at $t'=-10Z'_{R}$, $0$, and $10Z'_{R}$. The right and left edges of the focus at $t'=-10Z'_{R}$ and $10Z'_{R}$ are clipped by the front and rear edges of the pulse, respectively, leading to the asymmetric profile in $z'$.}
\label{fig:5}
\end{figure}

Figure \ref{fig:5} displays the longitudinal Poynting vector of a pulsed solution with $\beta_I = 0.5$, $n=\ell=0$, $\kappa'_0 w_0 = 20$, and $T=10Z_R'$. The moving focus maintains a near-constant profile and maximum over 10 Rayleigh ranges ($L = 10Z_R'$). The black dashed lines in Fig. \ref{fig:5}(a) mark the full width at half maximum boundary of the pulse $\hat{g}_{00}$, which travels at the speed of light. The pulse propagates from left to right, but only has an appreciable Poynting vector in the vicinity of the moving focus. Consistent with the super-Gaussian profile of the pulse and in contrast to Fig. \ref{fig:2}(g), the maximum of the Poynting vector increases and then decreases as it moves through the focal region. At $t' = -10Z'_R$ and $10Z'_R$, the boundary of the pulse encroaches on the focus from behind and ahead, respectively, causing the asymmetry in the longitudinal profile observed in Figs. \ref{fig:5}(b) and \ref{fig:5}(d).

\begin{figure}[H]
\centering\includegraphics[width=3in]{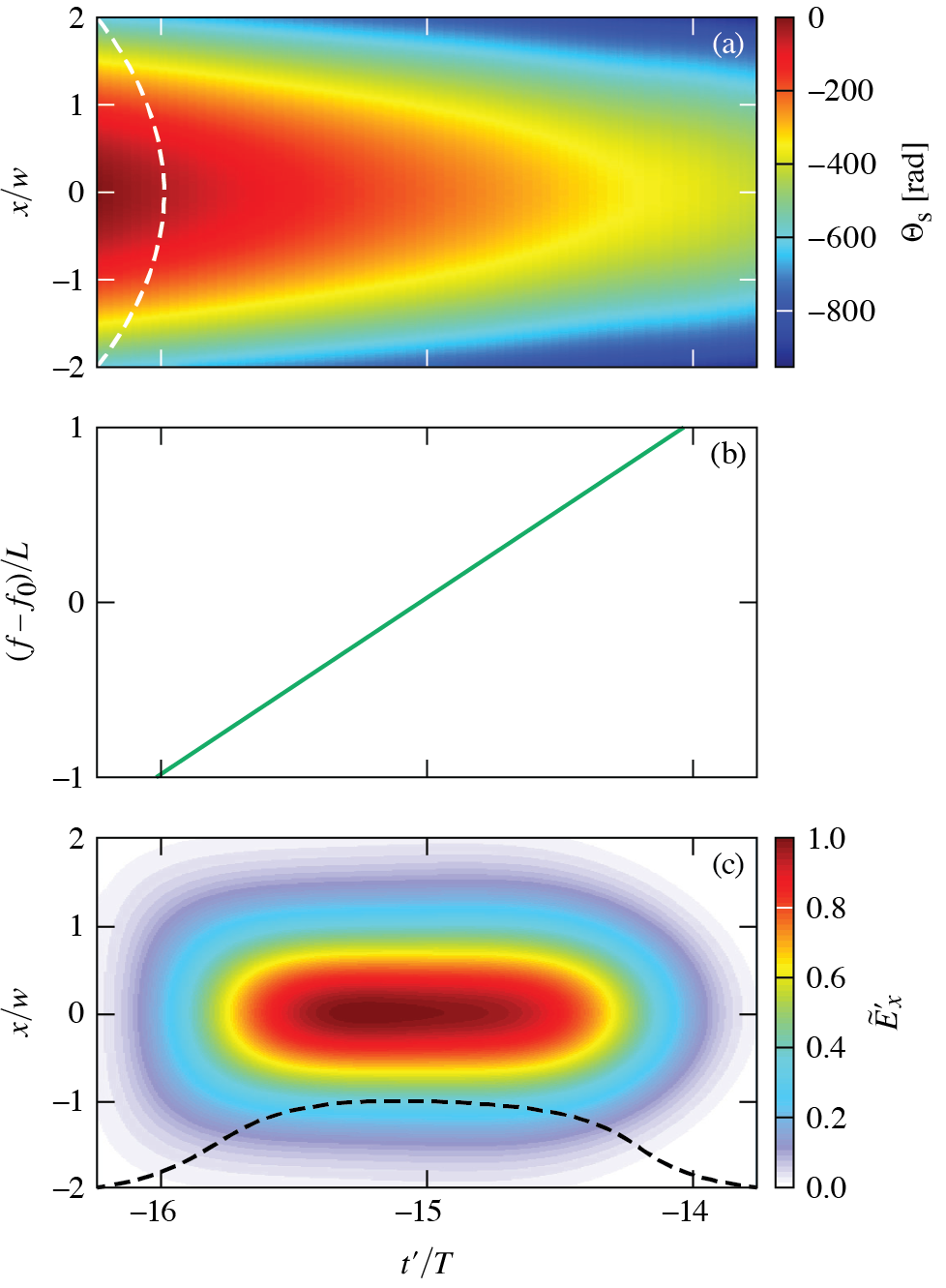}
\caption{Propagating the fields backward to $z' = -150 Z'_R$ provides (a) the slow phase $\Theta_s$, (b) the corresponding time-dependent focal length $f$, and (c) amplitude required to create a flying focus pulse with $\beta_I = 0.5$, $n = \ell = 0$, $\kappa'_0 w_0 = 20$, and $L = 10 Z'_R$ (the same parameters as in Fig. \ref{fig:4}). In (a) and (c) the dashed lines illustrate the parabolic shape of $\Theta_s$ and the on-axis amplitude, respectively. The transverse dimension is normalized to the spot size of a standard Gaussian beam at the same location: $w=w_0[1+(z'/Z_R')^2]^{1/2}$. More generally, a flying focus can be created by using a lens with focal length that depends linearly on time: $f(t') = f_0 + f_1 t'$, where $f_1$ is given by Eq. \eqref{eq:tslope}.}
\label{fig:6}
\end{figure}

The pulsed electromagnetic fields can be propagated backward in space to determine the amplitude and phase that an optical assembly must produce to realize these fields in an experiment. When discussing the phase, it is convenient to define a ``slow'' phase $\Theta_s$ which excludes the contribution from the carrier frequency, i.e., $\Theta_s \equiv \Theta + \kappa'_0 t'$, where $\Theta$ is the total phase. Figure \ref{fig:6}(a) shows the slow phase of the transverse electric field at $z'=-150Z_R'$ for the same parameters of Fig. \ref{fig:5}. The slow phase decreases linearly in time and has a transverse profile that is nearly parabolic. This is equivalent to the phase imparted by a lens with a focal length that depends linearly on time, which is consistent with Eq. \eqref{eq:duration} and the flying focus X. The time-dependent focal length $f(t')$ can be extracted from the phase by equating $\Theta_s = -\kappa_0'\rho^2/2f(t')$ [Fig. \ref{fig:6}(b)]. In general, one can show that $f(t') \approx f_0 + f_1 t'$, where
\begin{equation}\label{eq:tslope}
    f_1 = \frac{L}{T} =  \, \Bigr | \frac{\beta_I}{1-\beta_I} \Bigl |.
\end{equation}
Higher-order contributions to $f(t')$, e.g., $f_2t'^2$, drop as $1/z'^2$ or faster and thus are negligible in the near-field of the optical assembly. 

The amplitude of the transverse electric field is plotted in Fig. \ref{fig:6}(c). The transverse profile is nearly Gaussian with small deviations due to non-paraxiality. The temporal profile is nearly super-Gaussian, consistent with $\hat{g}_{n\ell}(t) \propto \exp[-(t/T)^4]$, but has an observable taper at earlier times. The later time slices of the pulse are farther from their focus and therefore have slightly larger spot sizes than the earlier time slices. In the case of a superluminal focus, the taper would be reversed, i.e., the earlier time slices would be closer to their focus and have smaller spot sizes. From conservation of power, the change in spot size throughout the pulse is accompanied by a change in amplitude. Specifically, 
\begin{equation}
    \frac{d|E_x'|}{dt'} \approx -\frac{\beta_I|1-\beta_I|Z_R'}{z'^2}.
\end{equation}
to leading order in $Z'_R/z'$. Far from the central focus $z'=0$, the variation in amplitude is negligible and can be ignored for experimental purposes.

\section{Comparison to paraxial solutions}\label{sec:PXvsHZ}

The exact expressions for the four-potential of a flying focus and the extension to pulsed solutions can be somewhat complicated, especially when considering higher-order radial or azimuthal modes. In many cases of interest, the bandwidth is narrow ($\kappa'_0 T \gg 1$), the spot size is much larger than the wavelength ($\kappa'_0 w_0 \gg 1$), and the vector nature of the field is unimportant. In these cases, paraxial solutions provide a simpler alternative to the full solutions. However, the reliability and accuracy of the paraxial solutions can only be determined through comparison to the exact solutions. This section presents such a comparison. 

Exact modal solutions for the four-potential in the paraxial approximation can be obtained without invoking the CSPM or a Lorentz transformation. For consistency with the Hertz vector formulation in the previous sections, consider the wave equation for the transverse vector potential in the Lorenz gauge:
\begin{equation}\label{eq:waveAperp}
    \left( \nabla_\perp^2+\partialDsq{z'} - \partialDsq{t'} \right)\textbf{A}'_{\perp}(\bvec{x}',t') = 0.
\end{equation}
Note that all quantities in Eq. \eqref{eq:waveAperp} are written in the laboratory frame. Upon performing the Galilean change of variables $\xi' = z'- \beta_I t'$ and $\eta' = z' -  t'$, the modal solution for $\textbf{A}'_{\perp}$ can be expressed as $\textbf{A}'_{\perp}(\bvec{x}',t') = \tfrac{1}{2}a'_{\perp}(\bvec{x}_\perp,\xi')e^{i \kappa' \eta'}\unitvec{e} + \text{c.c.}$, where the envelope $a'_{\perp}$ satisfies
\begin{equation} \label{eq:coordWave}  
    \left[ \nabla^2_\perp +(1-\beta_I^2)\partialDsq{\xi'} +2i\kappa'(1-\beta_I)\partialD{\xi'} \right]a'_{\perp}(\bvec{x}_{\perp},\xi')=0.
\end{equation}
With a solution to Eq. \eqref{eq:coordWave}, the remaining components of the four-potential can be calculated from $[2i\kappa' + (1+\beta_I)\partial_{\xi'}]A'_z = -\nabla_{\perp} \cdot \textbf{A}'_{\perp}$ and $\Phi' = -A'_z$. 

The closed-form solutions to Eq. \eqref{eq:coordWave} are identical to those described in Secs. \ref{sec:solutions} and \ref{sec:ExpSols}. While it is not clear how to arrive at these solutions directly from Eq. \eqref{eq:coordWave}, closed-form solutions can be obtained within the paraxial approximation. Specifically, Eq.  \eqref{eq:coordWave} reduces to the paraxial wave equation
\begin{equation} \label{eq:paraxialWave}  
    \left[ \nabla^2_\perp +2i\kappa'(1-\beta_I)\partialD{\xi'} \right]a'_{\perp}(\bvec{x}_{\perp},\xi')\approx 0
\end{equation}
when $|\kappa'(1-\beta_I)\partial_{\xi'}|\gg|(1-\beta_I^2)\partial^2_{\xi'}|$. Using the scaling $\kappa'(1-\beta_I)\partial_{\xi'} \sim \nabla_{\perp}^2 \sim w_0^{-2}$ evident in Eq. \eqref{eq:coordWave}, this condition can be reexpressed as
\begin{equation}\label{eq:paraxScal} 
    \kappa'^2 w_0^2\Bigl|\frac{1-\beta_I}{1+\beta_I}\Bigl| \gg 1.
\end{equation}
Inequality \eqref{eq:paraxScal} reveals that the paraxial approximation is accurate when $\beta_I \approx -1$ or when the variations of $a'_{\perp}$ with respect to $\xi'$ are slow compared to the frequency. Under these conditions, the solutions to Eq. \eqref{eq:coordWave} can be approximated as Laguerre--Gaussian modes ($\mathrm{LG}_{\ell p}$), such that
\begin{equation} \label{paraX}
\begin{aligned}
    A^{\prime P}_{x}(&\bvec{x}_{\perp},\eta',\xi') = 
    \frac{a_0w_0}{2w(\xi')}\left[ \frac{\sqrt{2} \rho}{w(\xi')}\right]^{|\ell|} \text{L}^{|\ell|}_p\left[\frac{2\rho^2}{w^2(\xi')}\right] \text{exp}\Bigg[i\kappa'\eta' - \Bigl(1-i\frac{\xi'}{\xi_0'}\Bigr)\frac{\rho^2}{w^2(\xi')}\\ 
    & \hspace{25pt}+i \ell \theta -i(2p+|\ell|+1) \text{arctan}\frac{\xi'}{\xi_0 '}\Bigg]+\mathrm{c.c.},\\
    \end{aligned}
\end{equation} 
where $\mathrm{L}^{|\ell|}_p$ is a generalized Laguerre polynomial, $w(\xi') = w_0 [1+(\xi'/\xi_0')^2]^{1/2}$, $\xi_0' = |1-\beta_I|Z_R'$, $\unitvec{e}=\unitvec{x}$ has been chosen for the polarization vector, and the superscript $P$ distinguishes paraxial solutions $A^{\prime P}_{x}$ from exact solutions $A^{\prime}_{x}$.

In the special case of $\beta_I = -1$, Eq. \eqref{eq:coordWave} is identical to Eq. \eqref{eq:paraxialWave}, and the Laguerre--Gaussian modes are exact solutions \cite{formanek2022radiation}. For $p = n = \ell = 0$, one can take the limit of the solutions in Sec. \ref{sec:ExpSols} as $\beta_I \rightarrow -1$ from above and below to show that $A^{\prime}_{x} = A^{\prime P}_{x}$. However, for $n>0$ or $|\ell|>0$, there is not a one-to-one mapping between the exact solutions generated from Eq. \eqref{eq:Sols} and the Laguerre--Gaussian modes. This can be readily verified by noting that Eq. \eqref{paraX} places no constraint on the integer values that $p$ and $\ell$ can take, whereas in Eq. \eqref{eq:Sols}, $|\ell| \leq n$. When $n>0$ or $|\ell|>0$, the exact solutions generated from Eq. \eqref{eq:Sols} are a superposition of multiple Lagerre-Gaussian modes with the same $\ell$ value. When $\beta_I = 1$, the Galilean coordinates become degenerate, i.e., $\xi' = \eta' = z'-t'$, thus, despite the equality of Eqs.  \eqref{eq:coordWave} and \eqref{eq:paraxialWave}, neither equation is valid. 

In the more general case of $|\beta_I| \neq 1$, the Laguerre--Gaussian modes are only approximate solutions. For $n = \ell = 0$, the exact solutions derived in Secs. \ref{sec:solutions} 
and \ref{sec:ExpSols} approach the $\mathrm{LG}_{00}$ mode in the paraxial limit. For $n>0$, the exact solutions approach a superposition of multiple LG modes with the same $\ell$ value. The one-to-one correspondence when $p=n=\ell=0$ allows for direct comparison of the exact and paraxial solutions.

The similarity of the exact and paraxial solutions can be quantified using the projection integral:
\begin{equation} \label{eq:proj}
\Gamma(\beta_I,\kappa'w_0)=\frac{\int  \langle A_{x}^{\prime}A_{x}^{\prime P}\rangle d\bvec{x}_\perp}{\sqrt{\int \langle A_{x}^{\prime}A_{x}^{\prime}\rangle d\bvec{x}_\perp}\sqrt{\int \langle A_{x}^{\prime P}A_{x}^{\prime P}\rangle d \bvec{x}_\perp}},
\end{equation} 
where $\langle \rangle$ denotes an average over the rapidly varying phase and $A^{\prime}_{x}$ and  $A^{\prime P}_{x}$ are evaluated in the focal plane $z' = \beta_I t'$. As defined, $\Gamma$ depends only on $\beta_I$ and $\kappa'w_0$. The values of $\Gamma \in [0,1]$. A value of $\Gamma = 1$ indicates that the vector potentials are identical, while a value of $\Gamma = 0$ indicates that they are orthogonal. The quantity $\langle A^{\prime}_{x} A^{\prime}_{x} \rangle$ was chosen for the projection because it is Lorentz invariant (recall from Sec. \ref{sec:solutions} that $A^{\mu}A_{\mu} = A_\perp^2$). 

\begin{figure}[H]
\centering\includegraphics[width=3in]{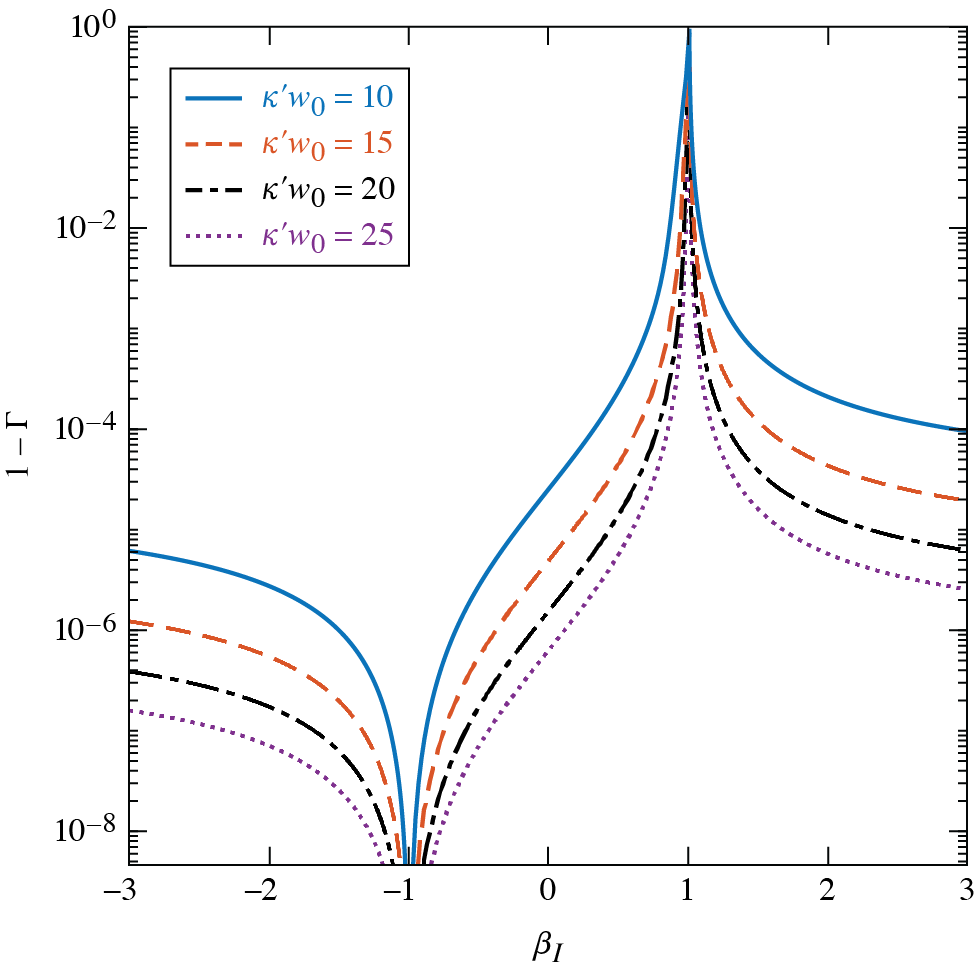}
\caption{One minus the projection integral [Eq. \eqref{eq:proj}] as a function of the focal velocity $\beta_I$ for different values of spot size $w_0$. The paraxial solutions are an excellent approximation to the exact solutions everywhere except for a small interval around $\beta_I = 1$.}
\label{fig:7}
\end{figure}

Figure \ref{fig:7} displays $1-\Gamma$ as a function of $\beta_I$ for several $\kappa'w_0$ values.  Consistent with inequality \eqref{eq:paraxScal}, the paraxial solution provides an excellent approximation to the exact solution everywhere except for a small interval around $\beta_I = 1$. The width of this interval narrows as $\kappa'w_0$ increases, which reflects the departure from paraxiality when the field is more tightly focused. As expected, $\Gamma \rightarrow 1$ as $\beta_I \rightarrow -1$.

\section{Summary and Conclusions}\label{sec:Conclusions}

The flying focus belongs to a broader class of optical techniques for controlling the space--time structure of laser pulses that also includes laser smoothing \cite{LEHMBERG198327,skupsky1989improved}, light sheets \cite{Kondakci2017,kondakci2019optical,yessenov2019classification}, and spatiotemporal optical vortices \cite{hancock2019free,chong2020generation,hancock2021stov}. Each of these has unique features that can lead to a deeper understanding of the fundamental properties of light or advance laser-based applications. The flying focus provides a cylindrically symmetric, programmable-velocity intensity peak that can travel distances far greater than a Rayleigh range while maintaining a near-constant profile \cite{sainte2017controlling,froula2018spatiotemporal,palastro2020dephasingless,simpson2022spatiotemporal}. These particular features can enable or enhance the broad range of applications that requires velocity matching and a high-intensity interaction over an extended distance. Assessing the extent to which a flying focus can improve these applications requires an accurate description of the electromagnetic fields.

Motivated by this requirement, this paper has presented a theoretical method for obtaining exact, closed form solutions for the electromagnetic fields of a constant-velocity flying focus. The method consists of three steps. In the first step, one finds multipole spherical or hyperbolic solutions to the wave equation that satisfy appropriate boundary conditions. These solutions are then converted into beam-like solutions using the complex source point method, i.e., by displacing one of the coordinates into its complex plane. The spherical and hyperbolic beam-like solutions have stationary foci in space and time, respectively. In the final step, the beam-like solutions are Lorentz boosted into a frame in which the foci appear to be moving. The spherical solutions produce subluminal foci, while the hyperbolic solutions produce superluminal foci. The method produces all six  components of the electromagnetic field, does not impose a paraxial approximation, and is generalized for arbitrary orbital angular momentum and polarization. 

Explicit expressions for the exact four-potential were provided to illustrate the structure of the solutions in select examples. The procedure for constructing pulsed solutions was then described. The pulsed solutions revealed that the peak of the Poynting vector travels at the focal velocity, whether it is positive or negative, while the sign of the Poynting vector always matches the direction of pulse propagation. Propagating the pulsed solutions backward in space demonstrated that the solutions describe a laser pulse focused by a lens with a focal length that depends linearly on time. As a result, these solutions may be generated in practice using a technique like the flying focus X  \cite{simpson2022spatiotemporal}. Alternatively, a time-ordered sequence of pulses each with a different focal length could be used as in the ASTRL technique \cite{pierce2022ASTRL}. Finally, it was shown that even in tight focusing geometries, approximate paraxial solutions for the flying focus \cite{ramsey2022nonlinear} can be accurate as long as the focal velocity is not too close to the speed of light. 
\medskip
\begin{acknowledgments}
The authors acknowledge Z. Barfield, J.L. Shaw, K. G. Miller, M.V. Ambat, J. Pigeon, and B. Barbosa for enthusiastic discussions.

This report was prepared as an account of work sponsored by an agency of the U.S. Government. Neither the U.S. Government nor any agency thereof, nor any of their employees, makes any warranty, express or implied, or assumes any legal liability or responsibility for the accuracy, completeness, or usefulness of any information, apparatus, product, or process disclosed, or represents that its use would not infringe privately owned rights. Reference herein to any specific commercial product, process, or service by trade name, trademark, manufacturer, or otherwise does not necessarily constitute or imply its endorsement, recommendation, or favoring by the U.S. Government or any agency thereof. The views and opinions of authors expressed herein do not necessarily state or reflect those of the U.S. Government or any agency thereof.

This material is based upon work supported by the Office of Fusion Energy Sciences under Award Number DE-SC0019135 and DE-SC00215057, the Department of Energy National Nuclear Security Administration under Award Number DE-NA0003856, the University of Rochester, and the New York State Energy Research and Development Authority.
\end{acknowledgments}


\begin{thebibliography}{54}%
\makeatletter
\providecommand \@ifxundefined [1]{%
 \@ifx{#1\undefined}
}%
\providecommand \@ifnum [1]{%
 \ifnum #1\expandafter \@firstoftwo
 \else \expandafter \@secondoftwo
 \fi
}%
\providecommand \@ifx [1]{%
 \ifx #1\expandafter \@firstoftwo
 \else \expandafter \@secondoftwo
 \fi
}%
\providecommand \natexlab [1]{#1}%
\providecommand \enquote  [1]{``#1''}%
\providecommand \bibnamefont  [1]{#1}%
\providecommand \bibfnamefont [1]{#1}%
\providecommand \citenamefont [1]{#1}%
\providecommand \href@noop [0]{\@secondoftwo}%
\providecommand \href [0]{\begingroup \@sanitize@url \@href}%
\providecommand \@href[1]{\@@startlink{#1}\@@href}%
\providecommand \@@href[1]{\endgroup#1\@@endlink}%
\providecommand \@sanitize@url [0]{\catcode `\\12\catcode `\$12\catcode
  `\&12\catcode `\#12\catcode `\^12\catcode `\_12\catcode `\%12\relax}%
\providecommand \@@startlink[1]{}%
\providecommand \@@endlink[0]{}%
\providecommand \url  [0]{\begingroup\@sanitize@url \@url }%
\providecommand \@url [1]{\endgroup\@href {#1}{\urlprefix }}%
\providecommand \urlprefix  [0]{URL }%
\providecommand \Eprint [0]{\href }%
\providecommand \doibase [0]{https://doi.org/}%
\providecommand \selectlanguage [0]{\@gobble}%
\providecommand \bibinfo  [0]{\@secondoftwo}%
\providecommand \bibfield  [0]{\@secondoftwo}%
\providecommand \translation [1]{[#1]}%
\providecommand \BibitemOpen [0]{}%
\providecommand \bibitemStop [0]{}%
\providecommand \bibitemNoStop [0]{.\EOS\space}%
\providecommand \EOS [0]{\spacefactor3000\relax}%
\providecommand \BibitemShut  [1]{\csname bibitem#1\endcsname}%
\let\auto@bib@innerbib\@empty
\bibitem [{\citenamefont {Sainte-Marie}\ \emph {et~al.}(2017)\citenamefont
  {Sainte-Marie}, \citenamefont {Gobert},\ and\ \citenamefont
  {Quere}}]{sainte2017controlling}%
  \BibitemOpen
  \bibfield  {author} {\bibinfo {author} {\bibfnamefont {A.}~\bibnamefont
  {Sainte-Marie}}, \bibinfo {author} {\bibfnamefont {O.}~\bibnamefont
  {Gobert}},\ and\ \bibinfo {author} {\bibfnamefont {F.}~\bibnamefont
  {Quere}},\ }\bibfield  {title} {\bibinfo {title} {Controlling the velocity of
  ultrashort light pulses in vacuum through spatio-temporal couplings},\
  }\href@noop {} {\bibfield  {journal} {\bibinfo  {journal} {Optica}\ }\textbf
  {\bibinfo {volume} {4}},\ \bibinfo {pages} {1298} (\bibinfo {year}
  {2017})}\BibitemShut {NoStop}%
\bibitem [{\citenamefont {Froula}\ \emph {et~al.}(2018)\citenamefont {Froula},
  \citenamefont {Turnbull}, \citenamefont {Davies}, \citenamefont {Kessler},
  \citenamefont {Haberberger}, \citenamefont {Palastro}, \citenamefont {Bahk},
  \citenamefont {Begishev}, \citenamefont {Boni}, \citenamefont {Bucht} \emph
  {et~al.}}]{froula2018spatiotemporal}%
  \BibitemOpen
  \bibfield  {author} {\bibinfo {author} {\bibfnamefont {D.~H.}\ \bibnamefont
  {Froula}}, \bibinfo {author} {\bibfnamefont {D.}~\bibnamefont {Turnbull}},
  \bibinfo {author} {\bibfnamefont {A.~S.}\ \bibnamefont {Davies}}, \bibinfo
  {author} {\bibfnamefont {T.~J.}\ \bibnamefont {Kessler}}, \bibinfo {author}
  {\bibfnamefont {D.}~\bibnamefont {Haberberger}}, \bibinfo {author}
  {\bibfnamefont {J.~P.}\ \bibnamefont {Palastro}}, \bibinfo {author}
  {\bibfnamefont {S.-W.}\ \bibnamefont {Bahk}}, \bibinfo {author}
  {\bibfnamefont {I.~A.}\ \bibnamefont {Begishev}}, \bibinfo {author}
  {\bibfnamefont {R.}~\bibnamefont {Boni}}, \bibinfo {author} {\bibfnamefont
  {S.}~\bibnamefont {Bucht}}, \emph {et~al.},\ }\bibfield  {title} {\bibinfo
  {title} {Spatiotemporal control of laser intensity},\ }\href@noop {}
  {\bibfield  {journal} {\bibinfo  {journal} {Nature photonics}\ }\textbf
  {\bibinfo {volume} {12}},\ \bibinfo {pages} {262} (\bibinfo {year}
  {2018})}\BibitemShut {NoStop}%
\bibitem [{\citenamefont {Simpson}\ \emph {et~al.}(2020)\citenamefont
  {Simpson}, \citenamefont {Ramsey}, \citenamefont {Franke}, \citenamefont
  {Vafaei-Najafabadi}, \citenamefont {Turnbull}, \citenamefont {Froula},\ and\
  \citenamefont {Palastro}}]{simpson2020nonlinear}%
  \BibitemOpen
  \bibfield  {author} {\bibinfo {author} {\bibfnamefont {T.~T.}\ \bibnamefont
  {Simpson}}, \bibinfo {author} {\bibfnamefont {D.}~\bibnamefont {Ramsey}},
  \bibinfo {author} {\bibfnamefont {P.}~\bibnamefont {Franke}}, \bibinfo
  {author} {\bibfnamefont {N.}~\bibnamefont {Vafaei-Najafabadi}}, \bibinfo
  {author} {\bibfnamefont {D.}~\bibnamefont {Turnbull}}, \bibinfo {author}
  {\bibfnamefont {D.~H.}\ \bibnamefont {Froula}},\ and\ \bibinfo {author}
  {\bibfnamefont {J.~P.}\ \bibnamefont {Palastro}},\ }\bibfield  {title}
  {\bibinfo {title} {Nonlinear spatiotemporal control of laser intensity},\
  }\href@noop {} {\bibfield  {journal} {\bibinfo  {journal} {Optics Express}\
  }\textbf {\bibinfo {volume} {28}},\ \bibinfo {pages} {38516} (\bibinfo {year}
  {2020})}\BibitemShut {NoStop}%
\bibitem [{\citenamefont {Simpson}\ \emph {et~al.}(2022)\citenamefont
  {Simpson}, \citenamefont {Ramsey}, \citenamefont {Franke}, \citenamefont
  {Weichman}, \citenamefont {Ambat}, \citenamefont {Turnbull}, \citenamefont
  {Froula},\ and\ \citenamefont {Palastro}}]{simpson2022spatiotemporal}%
  \BibitemOpen
  \bibfield  {author} {\bibinfo {author} {\bibfnamefont {T.~T.}\ \bibnamefont
  {Simpson}}, \bibinfo {author} {\bibfnamefont {D.}~\bibnamefont {Ramsey}},
  \bibinfo {author} {\bibfnamefont {P.}~\bibnamefont {Franke}}, \bibinfo
  {author} {\bibfnamefont {K.}~\bibnamefont {Weichman}}, \bibinfo {author}
  {\bibfnamefont {M.~V.}\ \bibnamefont {Ambat}}, \bibinfo {author}
  {\bibfnamefont {D.}~\bibnamefont {Turnbull}}, \bibinfo {author}
  {\bibfnamefont {D.~H.}\ \bibnamefont {Froula}},\ and\ \bibinfo {author}
  {\bibfnamefont {J.~P.}\ \bibnamefont {Palastro}},\ }\bibfield  {title}
  {\bibinfo {title} {Spatiotemporal control of laser intensity through
  cross-phase modulation},\ }\href@noop {} {\bibfield  {journal} {\bibinfo
  {journal} {Optics Express}\ }\textbf {\bibinfo {volume} {30}},\ \bibinfo
  {pages} {9878} (\bibinfo {year} {2022})}\BibitemShut {NoStop}%
\bibitem [{\citenamefont {Palastro}\ \emph {et~al.}(2020)\citenamefont
  {Palastro}, \citenamefont {Shaw}, \citenamefont {Franke}, \citenamefont
  {Ramsey}, \citenamefont {Simpson},\ and\ \citenamefont
  {Froula}}]{palastro2020dephasingless}%
  \BibitemOpen
  \bibfield  {author} {\bibinfo {author} {\bibfnamefont {J.}~\bibnamefont
  {Palastro}}, \bibinfo {author} {\bibfnamefont {J.}~\bibnamefont {Shaw}},
  \bibinfo {author} {\bibfnamefont {P.}~\bibnamefont {Franke}}, \bibinfo
  {author} {\bibfnamefont {D.}~\bibnamefont {Ramsey}}, \bibinfo {author}
  {\bibfnamefont {T.}~\bibnamefont {Simpson}},\ and\ \bibinfo {author}
  {\bibfnamefont {D.}~\bibnamefont {Froula}},\ }\bibfield  {title} {\bibinfo
  {title} {Dephasingless laser wakefield acceleration},\ }\href@noop {}
  {\bibfield  {journal} {\bibinfo  {journal} {Physical review letters}\
  }\textbf {\bibinfo {volume} {124}},\ \bibinfo {pages} {134802} (\bibinfo
  {year} {2020})}\BibitemShut {NoStop}%
\bibitem [{\citenamefont {Pierce}\ \emph {et~al.}(2022)\citenamefont {Pierce},
  \citenamefont {Palastro}, \citenamefont {Li}, \citenamefont {Malaca},
  \citenamefont {Ramsey}, \citenamefont {Vieira}, \citenamefont {Weichman},\
  and\ \citenamefont {Mori}}]{pierce2022ASTRL}%
  \BibitemOpen
  \bibfield  {author} {\bibinfo {author} {\bibfnamefont {J.~R.}\ \bibnamefont
  {Pierce}}, \bibinfo {author} {\bibfnamefont {J.~P.}\ \bibnamefont
  {Palastro}}, \bibinfo {author} {\bibfnamefont {F.}~\bibnamefont {Li}},
  \bibinfo {author} {\bibfnamefont {B.}~\bibnamefont {Malaca}}, \bibinfo
  {author} {\bibfnamefont {D.}~\bibnamefont {Ramsey}}, \bibinfo {author}
  {\bibfnamefont {J.}~\bibnamefont {Vieira}}, \bibinfo {author} {\bibfnamefont
  {K.}~\bibnamefont {Weichman}},\ and\ \bibinfo {author} {\bibfnamefont
  {W.~B.}\ \bibnamefont {Mori}},\ }\href
  {https://doi.org/10.48550/ARXIV.2207.13849} {\bibinfo {title} {Arbitrarily
  structured laser pulses}} (\bibinfo {year} {2022})\BibitemShut {NoStop}%
\bibitem [{\citenamefont {Smartsev}\ \emph {et~al.}(2019)\citenamefont
  {Smartsev}, \citenamefont {Caizergues}, \citenamefont {Oubrerie},
  \citenamefont {Gautier}, \citenamefont {Goddet}, \citenamefont {Tafzi},
  \citenamefont {Phuoc}, \citenamefont {Malka},\ and\ \citenamefont
  {Thaury}}]{smartsev2019axiparabola}%
  \BibitemOpen
  \bibfield  {author} {\bibinfo {author} {\bibfnamefont {S.}~\bibnamefont
  {Smartsev}}, \bibinfo {author} {\bibfnamefont {C.}~\bibnamefont
  {Caizergues}}, \bibinfo {author} {\bibfnamefont {K.}~\bibnamefont
  {Oubrerie}}, \bibinfo {author} {\bibfnamefont {J.}~\bibnamefont {Gautier}},
  \bibinfo {author} {\bibfnamefont {J.-P.}\ \bibnamefont {Goddet}}, \bibinfo
  {author} {\bibfnamefont {A.}~\bibnamefont {Tafzi}}, \bibinfo {author}
  {\bibfnamefont {K.~T.}\ \bibnamefont {Phuoc}}, \bibinfo {author}
  {\bibfnamefont {V.}~\bibnamefont {Malka}},\ and\ \bibinfo {author}
  {\bibfnamefont {C.}~\bibnamefont {Thaury}},\ }\bibfield  {title} {\bibinfo
  {title} {Axiparabola: a long-focal-depth, high-resolution mirror for
  broadband high-intensity lasers},\ }\href@noop {} {\bibfield  {journal}
  {\bibinfo  {journal} {Optics letters}\ }\textbf {\bibinfo {volume} {44}},\
  \bibinfo {pages} {3414} (\bibinfo {year} {2019})}\BibitemShut {NoStop}%
\bibitem [{\citenamefont {Debus}\ \emph {et~al.}(2019)\citenamefont {Debus},
  \citenamefont {Pausch}, \citenamefont {Huebl}, \citenamefont {Steiniger},
  \citenamefont {Widera}, \citenamefont {Cowan}, \citenamefont {Schramm},\ and\
  \citenamefont {Bussmann}}]{Debus2019}%
  \BibitemOpen
  \bibfield  {author} {\bibinfo {author} {\bibfnamefont {A.}~\bibnamefont
  {Debus}}, \bibinfo {author} {\bibfnamefont {R.}~\bibnamefont {Pausch}},
  \bibinfo {author} {\bibfnamefont {A.}~\bibnamefont {Huebl}}, \bibinfo
  {author} {\bibfnamefont {K.}~\bibnamefont {Steiniger}}, \bibinfo {author}
  {\bibfnamefont {R.}~\bibnamefont {Widera}}, \bibinfo {author} {\bibfnamefont
  {T.~E.}\ \bibnamefont {Cowan}}, \bibinfo {author} {\bibfnamefont
  {U.}~\bibnamefont {Schramm}},\ and\ \bibinfo {author} {\bibfnamefont
  {M.}~\bibnamefont {Bussmann}},\ }\bibfield  {title} {\bibinfo {title}
  {Circumventing the dephasing and depletion limits of laser-wakefield
  acceleration},\ }\bibfield  {journal} {\bibinfo  {journal} {Physical Review
  X}\ }\textbf {\bibinfo {volume} {9}},\ \href
  {https://doi.org/10.1103/physrevx.9.031044} {10.1103/physrevx.9.031044}
  (\bibinfo {year} {2019})\BibitemShut {NoStop}%
\bibitem [{\citenamefont {Caizergues}\ \emph {et~al.}(2020)\citenamefont
  {Caizergues}, \citenamefont {Smartsev}, \citenamefont {Malka},\ and\
  \citenamefont {Thaury}}]{Caizergues2020}%
  \BibitemOpen
  \bibfield  {author} {\bibinfo {author} {\bibfnamefont {C.}~\bibnamefont
  {Caizergues}}, \bibinfo {author} {\bibfnamefont {S.}~\bibnamefont
  {Smartsev}}, \bibinfo {author} {\bibfnamefont {V.}~\bibnamefont {Malka}},\
  and\ \bibinfo {author} {\bibfnamefont {C.}~\bibnamefont {Thaury}},\
  }\bibfield  {title} {\bibinfo {title} {Phase-locked laser-wakefield electron
  acceleration},\ }\href {https://doi.org/10.1038/s41566-020-0657-2} {\bibfield
   {journal} {\bibinfo  {journal} {Nature Photonics}\ }\textbf {\bibinfo
  {volume} {14}},\ \bibinfo {pages} {475} (\bibinfo {year} {2020})}\BibitemShut
  {NoStop}%
\bibitem [{\citenamefont {Palastro}\ \emph {et~al.}(2021)\citenamefont
  {Palastro}, \citenamefont {Malaca}, \citenamefont {Vieira}, \citenamefont
  {Ramsey}, \citenamefont {Simpson}, \citenamefont {Franke}, \citenamefont
  {Shaw},\ and\ \citenamefont {Froula}}]{palastro2021laser}%
  \BibitemOpen
  \bibfield  {author} {\bibinfo {author} {\bibfnamefont {J.}~\bibnamefont
  {Palastro}}, \bibinfo {author} {\bibfnamefont {B.}~\bibnamefont {Malaca}},
  \bibinfo {author} {\bibfnamefont {J.}~\bibnamefont {Vieira}}, \bibinfo
  {author} {\bibfnamefont {D.}~\bibnamefont {Ramsey}}, \bibinfo {author}
  {\bibfnamefont {T.}~\bibnamefont {Simpson}}, \bibinfo {author} {\bibfnamefont
  {P.}~\bibnamefont {Franke}}, \bibinfo {author} {\bibfnamefont
  {J.}~\bibnamefont {Shaw}},\ and\ \bibinfo {author} {\bibfnamefont
  {D.}~\bibnamefont {Froula}},\ }\bibfield  {title} {\bibinfo {title}
  {Laser-plasma acceleration beyond wave breaking},\ }\href@noop {} {\bibfield
  {journal} {\bibinfo  {journal} {Physics of Plasmas}\ }\textbf {\bibinfo
  {volume} {28}},\ \bibinfo {pages} {013109} (\bibinfo {year}
  {2021})}\BibitemShut {NoStop}%
\bibitem [{\citenamefont {Franke}\ \emph {et~al.}(2021)\citenamefont {Franke},
  \citenamefont {Ramsey}, \citenamefont {Simpson}, \citenamefont {Turnbull},
  \citenamefont {Froula},\ and\ \citenamefont {Palastro}}]{franke2021optical}%
  \BibitemOpen
  \bibfield  {author} {\bibinfo {author} {\bibfnamefont {P.}~\bibnamefont
  {Franke}}, \bibinfo {author} {\bibfnamefont {D.}~\bibnamefont {Ramsey}},
  \bibinfo {author} {\bibfnamefont {T.~T.}\ \bibnamefont {Simpson}}, \bibinfo
  {author} {\bibfnamefont {D.}~\bibnamefont {Turnbull}}, \bibinfo {author}
  {\bibfnamefont {D.}~\bibnamefont {Froula}},\ and\ \bibinfo {author}
  {\bibfnamefont {J.}~\bibnamefont {Palastro}},\ }\bibfield  {title} {\bibinfo
  {title} {Optical shock-enhanced self-photon acceleration},\ }\href@noop {}
  {\bibfield  {journal} {\bibinfo  {journal} {Physical Review A}\ }\textbf
  {\bibinfo {volume} {104}},\ \bibinfo {pages} {043520} (\bibinfo {year}
  {2021})}\BibitemShut {NoStop}%
\bibitem [{\citenamefont {Palastro}\ \emph {et~al.}(2018)\citenamefont
  {Palastro}, \citenamefont {Turnbull}, \citenamefont {Bahk}, \citenamefont
  {Follett}, \citenamefont {Shaw}, \citenamefont {Haberberger}, \citenamefont
  {Bromage},\ and\ \citenamefont {Froula}}]{palastro2018ionization}%
  \BibitemOpen
  \bibfield  {author} {\bibinfo {author} {\bibfnamefont {J.~P.}\ \bibnamefont
  {Palastro}}, \bibinfo {author} {\bibfnamefont {D.}~\bibnamefont {Turnbull}},
  \bibinfo {author} {\bibfnamefont {S.-W.}\ \bibnamefont {Bahk}}, \bibinfo
  {author} {\bibfnamefont {R.~K.}\ \bibnamefont {Follett}}, \bibinfo {author}
  {\bibfnamefont {J.~L.}\ \bibnamefont {Shaw}}, \bibinfo {author}
  {\bibfnamefont {D.}~\bibnamefont {Haberberger}}, \bibinfo {author}
  {\bibfnamefont {J.}~\bibnamefont {Bromage}},\ and\ \bibinfo {author}
  {\bibfnamefont {D.~H.}\ \bibnamefont {Froula}},\ }\bibfield  {title}
  {\bibinfo {title} {Ionization waves of arbitrary velocity driven by a flying
  focus},\ }\href@noop {} {\bibfield  {journal} {\bibinfo  {journal} {Physical
  Review A}\ }\textbf {\bibinfo {volume} {97}},\ \bibinfo {pages} {033835}
  (\bibinfo {year} {2018})}\BibitemShut {NoStop}%
\bibitem [{\citenamefont {Howard}\ \emph {et~al.}(2019)\citenamefont {Howard},
  \citenamefont {Turnbull}, \citenamefont {Davies}, \citenamefont {Franke},
  \citenamefont {Froula},\ and\ \citenamefont {Palastro}}]{howard2019photon}%
  \BibitemOpen
  \bibfield  {author} {\bibinfo {author} {\bibfnamefont {A.}~\bibnamefont
  {Howard}}, \bibinfo {author} {\bibfnamefont {D.}~\bibnamefont {Turnbull}},
  \bibinfo {author} {\bibfnamefont {A.}~\bibnamefont {Davies}}, \bibinfo
  {author} {\bibfnamefont {P.}~\bibnamefont {Franke}}, \bibinfo {author}
  {\bibfnamefont {D.}~\bibnamefont {Froula}},\ and\ \bibinfo {author}
  {\bibfnamefont {J.}~\bibnamefont {Palastro}},\ }\bibfield  {title} {\bibinfo
  {title} {Photon acceleration in a flying focus},\ }\href@noop {} {\bibfield
  {journal} {\bibinfo  {journal} {Physical review letters}\ }\textbf {\bibinfo
  {volume} {123}},\ \bibinfo {pages} {124801} (\bibinfo {year}
  {2019})}\BibitemShut {NoStop}%
\bibitem [{\citenamefont {Turnbull}\ \emph {et~al.}(2018)\citenamefont
  {Turnbull}, \citenamefont {Bucht}, \citenamefont {Davies}, \citenamefont
  {Haberberger}, \citenamefont {Kessler}, \citenamefont {Shaw},\ and\
  \citenamefont {Froula}}]{Turnbull2018}%
  \BibitemOpen
  \bibfield  {author} {\bibinfo {author} {\bibfnamefont {D.}~\bibnamefont
  {Turnbull}}, \bibinfo {author} {\bibfnamefont {S.}~\bibnamefont {Bucht}},
  \bibinfo {author} {\bibfnamefont {A.}~\bibnamefont {Davies}}, \bibinfo
  {author} {\bibfnamefont {D.}~\bibnamefont {Haberberger}}, \bibinfo {author}
  {\bibfnamefont {T.}~\bibnamefont {Kessler}}, \bibinfo {author} {\bibfnamefont
  {J.~L.}\ \bibnamefont {Shaw}},\ and\ \bibinfo {author} {\bibfnamefont
  {D.~H.}\ \bibnamefont {Froula}},\ }\bibfield  {title} {\bibinfo {title}
  {Raman amplification with a flying focus},\ }\href
  {https://doi.org/10.1103/PhysRevLett.120.024801} {\bibfield  {journal}
  {\bibinfo  {journal} {Phys. Rev. Lett.}\ }\textbf {\bibinfo {volume} {120}},\
  \bibinfo {pages} {024801} (\bibinfo {year} {2018})}\BibitemShut {NoStop}%
\bibitem [{\citenamefont {Di~Piazza}(2021)}]{di2021scattering}%
  \BibitemOpen
  \bibfield  {author} {\bibinfo {author} {\bibfnamefont {A.}~\bibnamefont
  {Di~Piazza}},\ }\bibfield  {title} {\bibinfo {title} {Unveiling the
  transverse formation length of nonlinear compton scattering},\ }\href
  {https://doi.org/10.1103/PhysRevA.103.012215} {\bibfield  {journal} {\bibinfo
   {journal} {Phys. Rev. A}\ }\textbf {\bibinfo {volume} {103}},\ \bibinfo
  {pages} {012215} (\bibinfo {year} {2021})}\BibitemShut {NoStop}%
\bibitem [{\citenamefont {Formanek}\ \emph {et~al.}(2022)\citenamefont
  {Formanek}, \citenamefont {Ramsey}, \citenamefont {Palastro},\ and\
  \citenamefont {Di~Piazza}}]{formanek2022radiation}%
  \BibitemOpen
  \bibfield  {author} {\bibinfo {author} {\bibfnamefont {M.}~\bibnamefont
  {Formanek}}, \bibinfo {author} {\bibfnamefont {D.}~\bibnamefont {Ramsey}},
  \bibinfo {author} {\bibfnamefont {J.}~\bibnamefont {Palastro}},\ and\
  \bibinfo {author} {\bibfnamefont {A.}~\bibnamefont {Di~Piazza}},\ }\bibfield
  {title} {\bibinfo {title} {Radiation reaction enhancement in flying focus
  pulses},\ }\href@noop {} {\bibfield  {journal} {\bibinfo  {journal} {Physical
  Review A}\ }\textbf {\bibinfo {volume} {105}},\ \bibinfo {pages} {L020203}
  (\bibinfo {year} {2022})}\BibitemShut {NoStop}%
\bibitem [{\citenamefont {Ramsey}\ \emph {et~al.}(2020)\citenamefont {Ramsey},
  \citenamefont {Franke}, \citenamefont {Simpson}, \citenamefont {Froula},\
  and\ \citenamefont {Palastro}}]{ramsey2020vacuum}%
  \BibitemOpen
  \bibfield  {author} {\bibinfo {author} {\bibfnamefont {D.}~\bibnamefont
  {Ramsey}}, \bibinfo {author} {\bibfnamefont {P.}~\bibnamefont {Franke}},
  \bibinfo {author} {\bibfnamefont {T.}~\bibnamefont {Simpson}}, \bibinfo
  {author} {\bibfnamefont {D.}~\bibnamefont {Froula}},\ and\ \bibinfo {author}
  {\bibfnamefont {J.}~\bibnamefont {Palastro}},\ }\bibfield  {title} {\bibinfo
  {title} {Vacuum acceleration of electrons in a dynamic laser pulse},\
  }\href@noop {} {\bibfield  {journal} {\bibinfo  {journal} {Physical Review
  E}\ }\textbf {\bibinfo {volume} {102}},\ \bibinfo {pages} {043207} (\bibinfo
  {year} {2020})}\BibitemShut {NoStop}%
\bibitem [{\citenamefont {Ramsey}\ \emph {et~al.}(2022)\citenamefont {Ramsey},
  \citenamefont {Malaca}, \citenamefont {Di~Piazza}, \citenamefont {Formanek},
  \citenamefont {Franke}, \citenamefont {Froula}, \citenamefont {Pardal},
  \citenamefont {Simpson}, \citenamefont {Vieira}, \citenamefont {Weichman}
  \emph {et~al.}}]{ramsey2022nonlinear}%
  \BibitemOpen
  \bibfield  {author} {\bibinfo {author} {\bibfnamefont {D.}~\bibnamefont
  {Ramsey}}, \bibinfo {author} {\bibfnamefont {B.}~\bibnamefont {Malaca}},
  \bibinfo {author} {\bibfnamefont {A.}~\bibnamefont {Di~Piazza}}, \bibinfo
  {author} {\bibfnamefont {M.}~\bibnamefont {Formanek}}, \bibinfo {author}
  {\bibfnamefont {P.}~\bibnamefont {Franke}}, \bibinfo {author} {\bibfnamefont
  {D.}~\bibnamefont {Froula}}, \bibinfo {author} {\bibfnamefont
  {M.}~\bibnamefont {Pardal}}, \bibinfo {author} {\bibfnamefont
  {T.}~\bibnamefont {Simpson}}, \bibinfo {author} {\bibfnamefont
  {J.}~\bibnamefont {Vieira}}, \bibinfo {author} {\bibfnamefont
  {K.}~\bibnamefont {Weichman}}, \emph {et~al.},\ }\bibfield  {title} {\bibinfo
  {title} {Nonlinear thomson scattering with ponderomotive control},\
  }\href@noop {} {\bibfield  {journal} {\bibinfo  {journal} {Physical Review
  E}\ }\textbf {\bibinfo {volume} {105}},\ \bibinfo {pages} {065201} (\bibinfo
  {year} {2022})}\BibitemShut {NoStop}%
\bibitem [{\citenamefont {Barton}\ \emph {et~al.}(1989)\citenamefont {Barton},
  \citenamefont {Alexander},\ and\ \citenamefont
  {Schaub}}]{barton1989theoretical}%
  \BibitemOpen
  \bibfield  {author} {\bibinfo {author} {\bibfnamefont {J.}~\bibnamefont
  {Barton}}, \bibinfo {author} {\bibfnamefont {D.}~\bibnamefont {Alexander}},\
  and\ \bibinfo {author} {\bibfnamefont {S.}~\bibnamefont {Schaub}},\
  }\bibfield  {title} {\bibinfo {title} {Theoretical determination of net
  radiation force and torque for a spherical particle illuminated by a focused
  laser beam},\ }\href@noop {} {\bibfield  {journal} {\bibinfo  {journal}
  {Journal of Applied Physics}\ }\textbf {\bibinfo {volume} {66}},\ \bibinfo
  {pages} {4594} (\bibinfo {year} {1989})}\BibitemShut {NoStop}%
\bibitem [{\citenamefont {Neuman}\ and\ \citenamefont
  {Block}(2004)}]{neuman2004optical}%
  \BibitemOpen
  \bibfield  {author} {\bibinfo {author} {\bibfnamefont {K.~C.}\ \bibnamefont
  {Neuman}}\ and\ \bibinfo {author} {\bibfnamefont {S.~M.}\ \bibnamefont
  {Block}},\ }\bibfield  {title} {\bibinfo {title} {Optical trapping},\
  }\href@noop {} {\bibfield  {journal} {\bibinfo  {journal} {Review of
  scientific instruments}\ }\textbf {\bibinfo {volume} {75}},\ \bibinfo {pages}
  {2787} (\bibinfo {year} {2004})}\BibitemShut {NoStop}%
\bibitem [{\citenamefont {Cicchitelli}\ \emph {et~al.}(1990)\citenamefont
  {Cicchitelli}, \citenamefont {Hora},\ and\ \citenamefont
  {Postle}}]{cicchitelli1990longitudinal}%
  \BibitemOpen
  \bibfield  {author} {\bibinfo {author} {\bibfnamefont {L.}~\bibnamefont
  {Cicchitelli}}, \bibinfo {author} {\bibfnamefont {H.}~\bibnamefont {Hora}},\
  and\ \bibinfo {author} {\bibfnamefont {R.}~\bibnamefont {Postle}},\
  }\bibfield  {title} {\bibinfo {title} {Longitudinal field components for
  laser beams in vacuum},\ }\href@noop {} {\bibfield  {journal} {\bibinfo
  {journal} {Physical Review A}\ }\textbf {\bibinfo {volume} {41}},\ \bibinfo
  {pages} {3727} (\bibinfo {year} {1990})}\BibitemShut {NoStop}%
\bibitem [{\citenamefont {Esarey}\ \emph
  {et~al.}(1995{\natexlab{a}})\citenamefont {Esarey}, \citenamefont {Sprangle},
  \citenamefont {Pilloff},\ and\ \citenamefont {Krall}}]{esarey1995theory}%
  \BibitemOpen
  \bibfield  {author} {\bibinfo {author} {\bibfnamefont {E.}~\bibnamefont
  {Esarey}}, \bibinfo {author} {\bibfnamefont {P.}~\bibnamefont {Sprangle}},
  \bibinfo {author} {\bibfnamefont {M.}~\bibnamefont {Pilloff}},\ and\ \bibinfo
  {author} {\bibfnamefont {J.}~\bibnamefont {Krall}},\ }\bibfield  {title}
  {\bibinfo {title} {Theory and group velocity of ultrashort, tightly focused
  laser pulses},\ }\href@noop {} {\bibfield  {journal} {\bibinfo  {journal}
  {JOSA B}\ }\textbf {\bibinfo {volume} {12}},\ \bibinfo {pages} {1695}
  (\bibinfo {year} {1995}{\natexlab{a}})}\BibitemShut {NoStop}%
\bibitem [{\citenamefont {Esarey}\ \emph
  {et~al.}(1995{\natexlab{b}})\citenamefont {Esarey}, \citenamefont
  {Sprangle},\ and\ \citenamefont {Krall}}]{esarey1995laser}%
  \BibitemOpen
  \bibfield  {author} {\bibinfo {author} {\bibfnamefont {E.}~\bibnamefont
  {Esarey}}, \bibinfo {author} {\bibfnamefont {P.}~\bibnamefont {Sprangle}},\
  and\ \bibinfo {author} {\bibfnamefont {J.}~\bibnamefont {Krall}},\ }\bibfield
   {title} {\bibinfo {title} {Laser acceleration of electrons in vacuum},\
  }\href@noop {} {\bibfield  {journal} {\bibinfo  {journal} {Physical Review
  E}\ }\textbf {\bibinfo {volume} {52}},\ \bibinfo {pages} {5443} (\bibinfo
  {year} {1995}{\natexlab{b}})}\BibitemShut {NoStop}%
\bibitem [{\citenamefont {Quesnel}\ and\ \citenamefont
  {Mora}(1998)}]{quesnel1998theory}%
  \BibitemOpen
  \bibfield  {author} {\bibinfo {author} {\bibfnamefont {B.}~\bibnamefont
  {Quesnel}}\ and\ \bibinfo {author} {\bibfnamefont {P.}~\bibnamefont {Mora}},\
  }\bibfield  {title} {\bibinfo {title} {Theory and simulation of the
  interaction of ultraintense laser pulses with electrons in vacuum},\
  }\href@noop {} {\bibfield  {journal} {\bibinfo  {journal} {Physical Review
  E}\ }\textbf {\bibinfo {volume} {58}},\ \bibinfo {pages} {3719} (\bibinfo
  {year} {1998})}\BibitemShut {NoStop}%
\bibitem [{\citenamefont {Hora}\ \emph {et~al.}(2000)\citenamefont {Hora},
  \citenamefont {Hoelss}, \citenamefont {Scheid}, \citenamefont {Wang},
  \citenamefont {Ho}, \citenamefont {Osman},\ and\ \citenamefont
  {Castillo}}]{hora2000principle}%
  \BibitemOpen
  \bibfield  {author} {\bibinfo {author} {\bibfnamefont {H.}~\bibnamefont
  {Hora}}, \bibinfo {author} {\bibfnamefont {M.}~\bibnamefont {Hoelss}},
  \bibinfo {author} {\bibfnamefont {W.}~\bibnamefont {Scheid}}, \bibinfo
  {author} {\bibfnamefont {J.}~\bibnamefont {Wang}}, \bibinfo {author}
  {\bibfnamefont {Y.}~\bibnamefont {Ho}}, \bibinfo {author} {\bibfnamefont
  {F.}~\bibnamefont {Osman}},\ and\ \bibinfo {author} {\bibfnamefont
  {R.}~\bibnamefont {Castillo}},\ }\bibfield  {title} {\bibinfo {title}
  {Principle of high accuracy for the nonlinear theory of the acceleration of
  electrons in a vacuum by lasers at relativistic intensities},\ }\href@noop {}
  {\bibfield  {journal} {\bibinfo  {journal} {Laser and Particle Beams}\
  }\textbf {\bibinfo {volume} {18}},\ \bibinfo {pages} {135} (\bibinfo {year}
  {2000})}\BibitemShut {NoStop}%
\bibitem [{\citenamefont {Lax}\ \emph {et~al.}(1975)\citenamefont {Lax},
  \citenamefont {Louisell},\ and\ \citenamefont {McKnight}}]{lax1975maxwell}%
  \BibitemOpen
  \bibfield  {author} {\bibinfo {author} {\bibfnamefont {M.}~\bibnamefont
  {Lax}}, \bibinfo {author} {\bibfnamefont {W.~H.}\ \bibnamefont {Louisell}},\
  and\ \bibinfo {author} {\bibfnamefont {W.~B.}\ \bibnamefont {McKnight}},\
  }\bibfield  {title} {\bibinfo {title} {From maxwell to paraxial wave
  optics},\ }\href@noop {} {\bibfield  {journal} {\bibinfo  {journal} {Physical
  Review A}\ }\textbf {\bibinfo {volume} {11}},\ \bibinfo {pages} {1365}
  (\bibinfo {year} {1975})}\BibitemShut {NoStop}%
\bibitem [{\citenamefont {Davis}(1979)}]{davis1979theory}%
  \BibitemOpen
  \bibfield  {author} {\bibinfo {author} {\bibfnamefont {L.}~\bibnamefont
  {Davis}},\ }\bibfield  {title} {\bibinfo {title} {Theory of electromagnetic
  beams},\ }\href@noop {} {\bibfield  {journal} {\bibinfo  {journal} {Physical
  Review A}\ }\textbf {\bibinfo {volume} {19}},\ \bibinfo {pages} {1177}
  (\bibinfo {year} {1979})}\BibitemShut {NoStop}%
\bibitem [{\citenamefont {Barton}\ and\ \citenamefont
  {Alexander}(1989)}]{barton1989fifth}%
  \BibitemOpen
  \bibfield  {author} {\bibinfo {author} {\bibfnamefont {J.~P.}\ \bibnamefont
  {Barton}}\ and\ \bibinfo {author} {\bibfnamefont {D.~R.}\ \bibnamefont
  {Alexander}},\ }\bibfield  {title} {\bibinfo {title} {Fifth-order corrected
  electromagnetic field components for a fundamental gaussian beam},\
  }\href@noop {} {\bibfield  {journal} {\bibinfo  {journal} {Journal of Applied
  Physics}\ }\textbf {\bibinfo {volume} {66}},\ \bibinfo {pages} {2800}
  (\bibinfo {year} {1989})}\BibitemShut {NoStop}%
\bibitem [{\citenamefont {Salamin}(2007)}]{salamin2007fields}%
  \BibitemOpen
  \bibfield  {author} {\bibinfo {author} {\bibfnamefont {Y.~I.}\ \bibnamefont
  {Salamin}},\ }\bibfield  {title} {\bibinfo {title} {Fields of a gaussian beam
  beyond the paraxial approximation},\ }\href@noop {} {\bibfield  {journal}
  {\bibinfo  {journal} {Applied Physics B}\ }\textbf {\bibinfo {volume} {86}},\
  \bibinfo {pages} {319} (\bibinfo {year} {2007})}\BibitemShut {NoStop}%
\bibitem [{\citenamefont {Agrawal}\ and\ \citenamefont
  {Pattanayak}(1979)}]{agrawal1979gaussian}%
  \BibitemOpen
  \bibfield  {author} {\bibinfo {author} {\bibfnamefont {G.~P.}\ \bibnamefont
  {Agrawal}}\ and\ \bibinfo {author} {\bibfnamefont {D.~N.}\ \bibnamefont
  {Pattanayak}},\ }\bibfield  {title} {\bibinfo {title} {Gaussian beam
  propagation beyond the paraxial approximation},\ }\href@noop {} {\bibfield
  {journal} {\bibinfo  {journal} {JOSA}\ }\textbf {\bibinfo {volume} {69}},\
  \bibinfo {pages} {575} (\bibinfo {year} {1979})}\BibitemShut {NoStop}%
\bibitem [{\citenamefont {Deschamps}(1971)}]{deschamps1971gaussian}%
  \BibitemOpen
  \bibfield  {author} {\bibinfo {author} {\bibfnamefont {G.~A.}\ \bibnamefont
  {Deschamps}},\ }\bibfield  {title} {\bibinfo {title} {Gaussian beam as a
  bundle of complex rays},\ }\href@noop {} {\bibfield  {journal} {\bibinfo
  {journal} {Electronics letters}\ }\textbf {\bibinfo {volume} {7}},\ \bibinfo
  {pages} {684} (\bibinfo {year} {1971})}\BibitemShut {NoStop}%
\bibitem [{\citenamefont {Shin}\ and\ \citenamefont
  {Felsen}(1977)}]{shin1977gaussian}%
  \BibitemOpen
  \bibfield  {author} {\bibinfo {author} {\bibfnamefont {S.~Y.}\ \bibnamefont
  {Shin}}\ and\ \bibinfo {author} {\bibfnamefont {L.}~\bibnamefont {Felsen}},\
  }\bibfield  {title} {\bibinfo {title} {Gaussian beam modes by multipoles with
  complex source points},\ }\href@noop {} {\bibfield  {journal} {\bibinfo
  {journal} {JOSA}\ }\textbf {\bibinfo {volume} {67}},\ \bibinfo {pages} {699}
  (\bibinfo {year} {1977})}\BibitemShut {NoStop}%
\bibitem [{\citenamefont {Zauderer}(1986)}]{zauderer1986complex}%
  \BibitemOpen
  \bibfield  {author} {\bibinfo {author} {\bibfnamefont {E.}~\bibnamefont
  {Zauderer}},\ }\bibfield  {title} {\bibinfo {title} {Complex argument
  hermite--gaussian and laguerre--gaussian beams},\ }\href@noop {} {\bibfield
  {journal} {\bibinfo  {journal} {JOSA A}\ }\textbf {\bibinfo {volume} {3}},\
  \bibinfo {pages} {465} (\bibinfo {year} {1986})}\BibitemShut {NoStop}%
\bibitem [{\citenamefont {Norris}(1986)}]{norris1986complex}%
  \BibitemOpen
  \bibfield  {author} {\bibinfo {author} {\bibfnamefont {A.}~\bibnamefont
  {Norris}},\ }\bibfield  {title} {\bibinfo {title} {Complex point-source
  representation of real point sources and the gaussian beam summation
  method},\ }\href@noop {} {\bibfield  {journal} {\bibinfo  {journal} {JOSA A}\
  }\textbf {\bibinfo {volume} {3}},\ \bibinfo {pages} {2005} (\bibinfo {year}
  {1986})}\BibitemShut {NoStop}%
\bibitem [{\citenamefont {Heyman}\ and\ \citenamefont
  {Felsen}(2001)}]{heyman2001gaussian}%
  \BibitemOpen
  \bibfield  {author} {\bibinfo {author} {\bibfnamefont {E.}~\bibnamefont
  {Heyman}}\ and\ \bibinfo {author} {\bibfnamefont {L.~B.}\ \bibnamefont
  {Felsen}},\ }\bibfield  {title} {\bibinfo {title} {Gaussian beam and
  pulsed-beam dynamics: complex-source and complex-spectrum formulations within
  and beyond paraxial asymptotics},\ }\href@noop {} {\bibfield  {journal}
  {\bibinfo  {journal} {JOSA A}\ }\textbf {\bibinfo {volume} {18}},\ \bibinfo
  {pages} {1588} (\bibinfo {year} {2001})}\BibitemShut {NoStop}%
\bibitem [{\citenamefont {Cullen}\ and\ \citenamefont
  {Yu}(1979)}]{cullen1979complex}%
  \BibitemOpen
  \bibfield  {author} {\bibinfo {author} {\bibfnamefont {A.~L.}\ \bibnamefont
  {Cullen}}\ and\ \bibinfo {author} {\bibfnamefont {P.}~\bibnamefont {Yu}},\
  }\bibfield  {title} {\bibinfo {title} {Complex source-point theory of the
  electromagnetic open resonator},\ }\href@noop {} {\bibfield  {journal}
  {\bibinfo  {journal} {Proceedings of the Royal Society of London. A.
  Mathematical and Physical Sciences}\ }\textbf {\bibinfo {volume} {366}},\
  \bibinfo {pages} {155} (\bibinfo {year} {1979})}\BibitemShut {NoStop}%
\bibitem [{\citenamefont {Sheppard}\ and\ \citenamefont
  {Saghafi}(1999)}]{sheppard1999electromagnetic}%
  \BibitemOpen
  \bibfield  {author} {\bibinfo {author} {\bibfnamefont {C.}~\bibnamefont
  {Sheppard}}\ and\ \bibinfo {author} {\bibfnamefont {S.}~\bibnamefont
  {Saghafi}},\ }\bibfield  {title} {\bibinfo {title} {Electromagnetic gaussian
  beams beyond the paraxial approximation},\ }\href@noop {} {\bibfield
  {journal} {\bibinfo  {journal} {JOSA A}\ }\textbf {\bibinfo {volume} {16}},\
  \bibinfo {pages} {1381} (\bibinfo {year} {1999})}\BibitemShut {NoStop}%
\bibitem [{\citenamefont {Mitri}(2013)}]{mitri2013quasi}%
  \BibitemOpen
  \bibfield  {author} {\bibinfo {author} {\bibfnamefont {F.}~\bibnamefont
  {Mitri}},\ }\bibfield  {title} {\bibinfo {title} {Quasi-gaussian
  electromagnetic beams},\ }\href@noop {} {\bibfield  {journal} {\bibinfo
  {journal} {Physical Review A}\ }\textbf {\bibinfo {volume} {87}},\ \bibinfo
  {pages} {035804} (\bibinfo {year} {2013})}\BibitemShut {NoStop}%
\bibitem [{\citenamefont {Saari}\ and\ \citenamefont
  {Besieris}(2020)}]{saari2020relativistic}%
  \BibitemOpen
  \bibfield  {author} {\bibinfo {author} {\bibfnamefont {P.}~\bibnamefont
  {Saari}}\ and\ \bibinfo {author} {\bibfnamefont {I.~M.}\ \bibnamefont
  {Besieris}},\ }\bibfield  {title} {\bibinfo {title} {Relativistic aberration
  and null doppler shift within the framework of superluminal and subluminal
  nondiffracting waves},\ }\href@noop {} {\bibfield  {journal} {\bibinfo
  {journal} {Journal of Physics Communications}\ }\textbf {\bibinfo {volume}
  {4}},\ \bibinfo {pages} {105011} (\bibinfo {year} {2020})}\BibitemShut
  {NoStop}%
\bibitem [{\citenamefont {Besieris}\ and\ \citenamefont
  {Saari}(2022)}]{besieris2022autofocusing}%
  \BibitemOpen
  \bibfield  {author} {\bibinfo {author} {\bibfnamefont {I.~M.}\ \bibnamefont
  {Besieris}}\ and\ \bibinfo {author} {\bibfnamefont {P.}~\bibnamefont
  {Saari}},\ }\bibfield  {title} {\bibinfo {title} {Autofocusing luminal and
  superluminal spatiotemporally localized waves},\ }\href@noop {} {\bibfield
  {journal} {\bibinfo  {journal} {JOSA A}\ }\textbf {\bibinfo {volume} {39}},\
  \bibinfo {pages} {1449} (\bibinfo {year} {2022})}\BibitemShut {NoStop}%
\bibitem [{\citenamefont {Yessenov}\ \emph {et~al.}(2022)\citenamefont
  {Yessenov}, \citenamefont {Hall}, \citenamefont {Schepler},\ and\
  \citenamefont {Abouraddy}}]{Yessenov22}%
  \BibitemOpen
  \bibfield  {author} {\bibinfo {author} {\bibfnamefont {M.}~\bibnamefont
  {Yessenov}}, \bibinfo {author} {\bibfnamefont {L.~A.}\ \bibnamefont {Hall}},
  \bibinfo {author} {\bibfnamefont {K.~L.}\ \bibnamefont {Schepler}},\ and\
  \bibinfo {author} {\bibfnamefont {A.~F.}\ \bibnamefont {Abouraddy}},\
  }\bibfield  {title} {\bibinfo {title} {Space-time wave packets},\ }\href
  {https://doi.org/10.1364/AOP.450016} {\bibfield  {journal} {\bibinfo
  {journal} {Adv. Opt. Photon.}\ }\textbf {\bibinfo {volume} {14}},\ \bibinfo
  {pages} {455} (\bibinfo {year} {2022})}\BibitemShut {NoStop}%
\bibitem [{\citenamefont {Longhi}(2004)}]{longhi2004gaussian}%
  \BibitemOpen
  \bibfield  {author} {\bibinfo {author} {\bibfnamefont {S.}~\bibnamefont
  {Longhi}},\ }\bibfield  {title} {\bibinfo {title} {Gaussian pulsed beams with
  arbitrary speed},\ }\href@noop {} {\bibfield  {journal} {\bibinfo  {journal}
  {Optics Express}\ }\textbf {\bibinfo {volume} {12}},\ \bibinfo {pages} {935}
  (\bibinfo {year} {2004})}\BibitemShut {NoStop}%
\bibitem [{\citenamefont {B{\'e}langer}(1986)}]{belanger1986lorentz}%
  \BibitemOpen
  \bibfield  {author} {\bibinfo {author} {\bibfnamefont {P.}~\bibnamefont
  {B{\'e}langer}},\ }\bibfield  {title} {\bibinfo {title} {Lorentz
  transformation of packetlike solutions of the homogeneous-wave equation},\
  }\href@noop {} {\bibfield  {journal} {\bibinfo  {journal} {JOSA A}\ }\textbf
  {\bibinfo {volume} {3}},\ \bibinfo {pages} {541} (\bibinfo {year}
  {1986})}\BibitemShut {NoStop}%
\bibitem [{\citenamefont {Jackson}(1999)}]{jackson_classical_1999}%
  \BibitemOpen
  \bibfield  {author} {\bibinfo {author} {\bibfnamefont {J.~D.}\ \bibnamefont
  {Jackson}},\ }\href {http://cdsweb.cern.ch/record/490457} {\emph {\bibinfo
  {title} {Classical electrodynamics}}},\ \bibinfo {edition} {3rd}\ ed.\
  (\bibinfo  {publisher} {Wiley},\ \bibinfo {address} {New York, {NY}},\
  \bibinfo {year} {1999})\BibitemShut {NoStop}%
\bibitem [{\citenamefont {Nisbet}(1955)}]{Nisbet1955Hertz}%
  \BibitemOpen
  \bibfield  {author} {\bibinfo {author} {\bibfnamefont {A.}~\bibnamefont
  {Nisbet}},\ }\bibfield  {title} {\bibinfo {title} {Hertzian electromagnetic
  potentials and associated gauge transformations},\ }\href
  {http://www.jstor.org/stable/99751} {\bibfield  {journal} {\bibinfo
  {journal} {Proceedings of the Royal Society of London. Series A, Mathematical
  and Physical Sciences}\ }\textbf {\bibinfo {volume} {231}},\ \bibinfo {pages}
  {250} (\bibinfo {year} {1955})}\BibitemShut {NoStop}%
\bibitem [{\citenamefont {Levy}\ \emph {et~al.}(2019)\citenamefont {Levy},
  \citenamefont {Silberberg},\ and\ \citenamefont
  {Davidson}}]{levy2019mathematics}%
  \BibitemOpen
  \bibfield  {author} {\bibinfo {author} {\bibfnamefont {U.}~\bibnamefont
  {Levy}}, \bibinfo {author} {\bibfnamefont {Y.}~\bibnamefont {Silberberg}},\
  and\ \bibinfo {author} {\bibfnamefont {N.}~\bibnamefont {Davidson}},\
  }\bibfield  {title} {\bibinfo {title} {Mathematics of vectorial gaussian
  beams},\ }\href@noop {} {\bibfield  {journal} {\bibinfo  {journal} {Advances
  in Optics and Photonics}\ }\textbf {\bibinfo {volume} {11}},\ \bibinfo
  {pages} {828} (\bibinfo {year} {2019})}\BibitemShut {NoStop}%
\bibitem [{\citenamefont {Lehmberg}\ and\ \citenamefont
  {Obenschain}(1983)}]{LEHMBERG198327}%
  \BibitemOpen
  \bibfield  {author} {\bibinfo {author} {\bibfnamefont {R.}~\bibnamefont
  {Lehmberg}}\ and\ \bibinfo {author} {\bibfnamefont {S.}~\bibnamefont
  {Obenschain}},\ }\bibfield  {title} {\bibinfo {title} {Use of induced spatial
  incoherence for uniform illumination of laser fusion targets},\ }\href
  {https://doi.org/https://doi.org/10.1016/0030-4018(83)90024-X} {\bibfield
  {journal} {\bibinfo  {journal} {Optics Communications}\ }\textbf {\bibinfo
  {volume} {46}},\ \bibinfo {pages} {27} (\bibinfo {year} {1983})}\BibitemShut
  {NoStop}%
\bibitem [{\citenamefont {Skupsky}\ \emph {et~al.}(1989)\citenamefont
  {Skupsky}, \citenamefont {Short}, \citenamefont {Kessler}, \citenamefont
  {Craxton}, \citenamefont {Letzring},\ and\ \citenamefont
  {Soures}}]{skupsky1989improved}%
  \BibitemOpen
  \bibfield  {author} {\bibinfo {author} {\bibfnamefont {S.}~\bibnamefont
  {Skupsky}}, \bibinfo {author} {\bibfnamefont {R.}~\bibnamefont {Short}},
  \bibinfo {author} {\bibfnamefont {T.}~\bibnamefont {Kessler}}, \bibinfo
  {author} {\bibfnamefont {R.}~\bibnamefont {Craxton}}, \bibinfo {author}
  {\bibfnamefont {S.}~\bibnamefont {Letzring}},\ and\ \bibinfo {author}
  {\bibfnamefont {J.}~\bibnamefont {Soures}},\ }\bibfield  {title} {\bibinfo
  {title} {Improved laser-beam uniformity using the angular dispersion of
  frequency-modulated light},\ }\href@noop {} {\bibfield  {journal} {\bibinfo
  {journal} {Journal of Applied Physics}\ }\textbf {\bibinfo {volume} {66}},\
  \bibinfo {pages} {3456} (\bibinfo {year} {1989})}\BibitemShut {NoStop}%
\bibitem [{\citenamefont {Kondakci}\ and\ \citenamefont
  {Abouraddy}(2017)}]{Kondakci2017}%
  \BibitemOpen
  \bibfield  {author} {\bibinfo {author} {\bibfnamefont {H.~E.}\ \bibnamefont
  {Kondakci}}\ and\ \bibinfo {author} {\bibfnamefont {A.~F.}\ \bibnamefont
  {Abouraddy}},\ }\bibfield  {title} {\bibinfo {title} {Diffraction-free
  space{\textendash}time light sheets},\ }\href
  {https://doi.org/10.1038/s41566-017-0028-9} {\bibfield  {journal} {\bibinfo
  {journal} {Nature Photonics}\ }\textbf {\bibinfo {volume} {11}},\ \bibinfo
  {pages} {733} (\bibinfo {year} {2017})}\BibitemShut {NoStop}%
\bibitem [{\citenamefont {Kondakci}\ and\ \citenamefont
  {Abouraddy}(2019)}]{kondakci2019optical}%
  \BibitemOpen
  \bibfield  {author} {\bibinfo {author} {\bibfnamefont {H.}~\bibnamefont
  {Kondakci}}\ and\ \bibinfo {author} {\bibfnamefont {A.~F.}\ \bibnamefont
  {Abouraddy}},\ }\bibfield  {title} {\bibinfo {title} {Optical space-time wave
  packets having arbitrary group velocities in free space},\ }\href@noop {}
  {\bibfield  {journal} {\bibinfo  {journal} {Nature communications}\ }\textbf
  {\bibinfo {volume} {10}},\ \bibinfo {pages} {1} (\bibinfo {year}
  {2019})}\BibitemShut {NoStop}%
\bibitem [{\citenamefont {Yessenov}\ \emph {et~al.}(2019)\citenamefont
  {Yessenov}, \citenamefont {Bhaduri}, \citenamefont {Kondakci},\ and\
  \citenamefont {Abouraddy}}]{yessenov2019classification}%
  \BibitemOpen
  \bibfield  {author} {\bibinfo {author} {\bibfnamefont {M.}~\bibnamefont
  {Yessenov}}, \bibinfo {author} {\bibfnamefont {B.}~\bibnamefont {Bhaduri}},
  \bibinfo {author} {\bibfnamefont {H.~E.}\ \bibnamefont {Kondakci}},\ and\
  \bibinfo {author} {\bibfnamefont {A.~F.}\ \bibnamefont {Abouraddy}},\
  }\bibfield  {title} {\bibinfo {title} {Classification of
  propagation-invariant space-time wave packets in free space: Theory and
  experiments},\ }\href@noop {} {\bibfield  {journal} {\bibinfo  {journal}
  {Physical Review A}\ }\textbf {\bibinfo {volume} {99}},\ \bibinfo {pages}
  {023856} (\bibinfo {year} {2019})}\BibitemShut {NoStop}%
\bibitem [{\citenamefont {Hancock}\ \emph {et~al.}(2019)\citenamefont
  {Hancock}, \citenamefont {Zahedpour}, \citenamefont {Goffin},\ and\
  \citenamefont {Milchberg}}]{hancock2019free}%
  \BibitemOpen
  \bibfield  {author} {\bibinfo {author} {\bibfnamefont {S.}~\bibnamefont
  {Hancock}}, \bibinfo {author} {\bibfnamefont {S.}~\bibnamefont {Zahedpour}},
  \bibinfo {author} {\bibfnamefont {A.}~\bibnamefont {Goffin}},\ and\ \bibinfo
  {author} {\bibfnamefont {H.}~\bibnamefont {Milchberg}},\ }\bibfield  {title}
  {\bibinfo {title} {Free-space propagation of spatiotemporal optical
  vortices},\ }\href@noop {} {\bibfield  {journal} {\bibinfo  {journal}
  {Optica}\ }\textbf {\bibinfo {volume} {6}},\ \bibinfo {pages} {1547}
  (\bibinfo {year} {2019})}\BibitemShut {NoStop}%
\bibitem [{\citenamefont {Chong}\ \emph {et~al.}(2020)\citenamefont {Chong},
  \citenamefont {Wan}, \citenamefont {Chen},\ and\ \citenamefont
  {Zhan}}]{chong2020generation}%
  \BibitemOpen
  \bibfield  {author} {\bibinfo {author} {\bibfnamefont {A.}~\bibnamefont
  {Chong}}, \bibinfo {author} {\bibfnamefont {C.}~\bibnamefont {Wan}}, \bibinfo
  {author} {\bibfnamefont {J.}~\bibnamefont {Chen}},\ and\ \bibinfo {author}
  {\bibfnamefont {Q.}~\bibnamefont {Zhan}},\ }\bibfield  {title} {\bibinfo
  {title} {Generation of spatiotemporal optical vortices with controllable
  transverse orbital angular momentum},\ }\href@noop {} {\bibfield  {journal}
  {\bibinfo  {journal} {Nature Photonics}\ }\textbf {\bibinfo {volume} {14}},\
  \bibinfo {pages} {350} (\bibinfo {year} {2020})}\BibitemShut {NoStop}%
\bibitem [{\citenamefont {Hancock}\ \emph {et~al.}(2021)\citenamefont
  {Hancock}, \citenamefont {Zahedpour},\ and\ \citenamefont
  {Milchberg}}]{hancock2021stov}%
  \BibitemOpen
  \bibfield  {author} {\bibinfo {author} {\bibfnamefont {S.~W.}\ \bibnamefont
  {Hancock}}, \bibinfo {author} {\bibfnamefont {S.}~\bibnamefont {Zahedpour}},\
  and\ \bibinfo {author} {\bibfnamefont {H.~M.}\ \bibnamefont {Milchberg}},\
  }\bibfield  {title} {\bibinfo {title} {Mode structure and orbital angular
  momentum of spatiotemporal optical vortex pulses},\ }\href
  {https://doi.org/10.1103/PhysRevLett.127.193901} {\bibfield  {journal}
  {\bibinfo  {journal} {Phys. Rev. Lett.}\ }\textbf {\bibinfo {volume} {127}},\
  \bibinfo {pages} {193901} (\bibinfo {year} {2021})}\BibitemShut {NoStop}%
\end{thebibliography}
%
\end{document}